\documentclass[twocolumn]{aastex631}

\usepackage{amsmath}
\usepackage{xspace} 
\usepackage{hyperref}
\usepackage{multirow}
\usepackage{xcolor}

\newcommand{\swift}{\textit{Swift}\xspace}
\newcommand{\nicer}{\textit{NICER}\xspace}

\newcommand{\xmm}{\textit{XMM-Newton}\xspace}
\newcommand{\hst}{\textit{HST}\xspace}
\newcommand{\pion}{\texttt{pion}\xspace}

\newcommand{\afhd}{AT2020afhd\xspace}
\newcommand{\wangetal}{Wang et al. (2025, submitted)\xspace}

\newcommand{\dstat}{$\Delta {\rm Cstat}$\xspace}
\newcommand{\pna}{PN101\xspace}
\newcommand{\pnb}{PN201\xspace}
\newcommand{\pnc}{PN301\xspace}

\begin{document}

\title{Delayed Launch of Ultrafast Outflows in the Tidal Disruption Event \afhd}

\correspondingauthor{Yanan Wang}
\email{wangyn@bao.ac.cn}

\author[0000-0001-9576-1870]{Zikun Lin}
\affiliation{Key Laboratory of Optical Astronomy, National Astronomical Observatories, Chinese Academy of Sciences, Beijing 100101, People’s Republic of China}
\affiliation{School of Astronomy and Space Sciences, University of Chinese Academy of Sciences, Beijing 100049, People’s Republic of China}

\author[0000-0003-3207-5237]{Yanan Wang}
\affiliation{Key Laboratory of Optical Astronomy, National Astronomical Observatories, Chinese Academy of Sciences, Beijing 100101, People’s Republic of China}

\author[0000-0002-0427-520X]{De-Fu Bu}
\affiliation{Shanghai Key Lab for Astrophysics, Shanghai Normal University, 100 Guilin Road, Shanghai 200234, People’s Republic of China}

\author[0000-0001-7557-9713]{Junjie Mao}
\affiliation{Department of Astronomy, Tsinghua University, Haidian DS 100084 Beijing, People’s Republic of China}

\author[0000-0002-2874-2706]{Jifeng Liu}
\affiliation{Key Laboratory of Optical Astronomy, National Astronomical Observatories, Chinese Academy of Sciences, Beijing 100101, People’s Republic of China}
\affiliation{School of Astronomy and Space Sciences, University of Chinese Academy of Sciences, Beijing 100049, People’s Republic of China}
\affiliation{Institute for Frontiers in Astronomy and Astrophysics, Beijing Normal University, Beijing 102206, People’s Republic of China}
\affiliation{New Cornerstone Science Laboratory, National Astronomical Observatories, Chinese Academy of Sciences, Beijing 100012, People’s Republic of China}

\begin{abstract}

We report the detection and characterization of ultrafast outflows (UFOs) in the X-ray spectra of the tidal disruption event (TDE) \afhd, based on observations from \nicer, \swift, and \xmm. Prominent blueshifted absorption features were detected exclusively during the intermediate phase of the event, occurring between days 172 and 212 within the first 300 days post-discovery. During this period, the UFO appeared no earlier than day 74, strengthened between days 172 and 194, and disappeared after day 215. 
This marks the first time that the full evolutionary sequence of X-ray outflows has been observed in a TDE.
Moreover, the outflows exhibited a dramatic deceleration from $\sim0.19$c to $\sim0.0097$c over a span of approximately 10\,days. Photoionization spectral analysis reveals an inverse correlation between outflow velocity and ionization parameter, in contradiction to the predictions from radiation pressure-driven wind. Eventually, we propose that the delayed onset of the outflows may result from an increase in the wind opening angle and/or metal enrichment, particularly iron and oxygen, during the disk formation phase.

\end{abstract}


\section{Introduction} \label{sec:intro}

When a star moves too close to a supermassive black hole (SMBH) and is torn apart by tidal forces, the resulting phenomenon is defined as a tidal disruption event \citep[TDE; see e.g.,][]{Hills1975,Rees1988Nat,Gezari2021}. Approximately half of the stellar debris becomes gravitationally bound to the SMBH \citep{Rees1988Nat,Phinney1989}, resulting in an accretion rate exceeding the Eddington limit at early times \citep[see e.g.,][]{Loeb1997}. It is widely accepted that super-Eddington accreting systems with strong radiation pressure could launch optically thick disk winds \citep[see e.g.,][]{Ohsuga2005,Strubbe2009,Jiang2014,McKinney2014,Metzger2016,Dai2018, Curd2019, Bu2022MNRAS, Thomsen2022, Qiao2025}. 

Outflows, identified by blueshifted absorption features, have been observed in some TDEs in the X-ray, ultraviolet (UV), or both wavelength.
In the X-ray band, such features have been observed in ASASSN-14li \citep{Miller2015,Kara2018}, 3XMM J152130.7+074916 \citep{Lin2015ApJ...811...43L}, ASASSN-20qc \citep{Kosec2023, Pasham2024}, AT2020ksf \citep{Wevers2023}, and AT2021ehb \citep{Xiang2024}. In the UV band, winds have been identified in PS1-11af \citep{Chornock2014}, ASASSN-14li \citep{Cenko2016}, PS16dtm \citep{Blanchard2017}, iPTF16fnl \citep{Brown2018}, iPTF15af \citep{Blagorodnova2019}, AT2018zr \citep{Hung2019}, AT2019qiz \citep{Hung2021}, and ASASSN-14ko \citep{Payne2023}.
In most of these events, the absorption features exhibit outflow velocities of $\gtrsim10^{4}$\,km\,s$^{-1}$, corresponding to ultrafast outflows (UFOs), whereas a few TDEs show slower winds with velocities of $\sim10^{2}$--$10^{4}$\,km\,s$^{-1}$ (e.g., \citealt{Chornock2014,Miller2015,Cenko2016,Kosec2023,Payne2023}).
Among these events, ASASSN-14li and ASASSN-20qc show multiple outflow components, including a UFO with a velocity of 0.2--0.3c \citep{Kara2018,Pasham2024} and slower components with velocities of 100--1000\,km\,s$^{-1}$ \citep{Miller2015,Cenko2016,Kosec2023}.

Although radiation pressure is often proposed as the primary driver of winds in TDEs (e.g., \citealt{Dai2018,Curd2019,Thomsen2022}), alternative mechanisms may also contribute significantly. For instance, magnetically driven winds have been proposed for the outflow observed in AT2021ehb \citep{Xiang2024}, while the slower components observed in ASASSN-20qc may be attributed to a thermal origin \citep{Kosec2023}.

Outflows may offer a solution to the ``missing energy problem" in TDEs \citep[see e.g.,][and references therein]{Lu2018ApJ}, where the observed radiative energy is significantly lower than the theoretically predicted energy released by the accretion of the disrupted stellar mass. This discrepancy may arise from other factors such as energy released at unobservable extreme UV wavelengths, absorption by the host galaxy, inefficient radiation due to photon trapping \citep[e.g.,][]{Loeb1997,Krolik2012}, mass directly fall onto the black hole due to high eccentricity of debris \citep{Svirski2017}, and partial disruptions \citep[e.g.,][]{Guillochon2013}. In addition, late-time UV observations reveal that the accretion disk can continue to emit a substantial fraction of its total energy in the UV band thousands of days after the disruption, deviating from the initially expected power-law decay \citep{vanVelzen2019}.
However, observational evidence to examine these factors remains limited. In particular, the detection of outflows (both jets and winds) is highly dependent on the system’s viewing angle. Moreover, the requirement for well-sampled, high signal-to-noise ratio (SNR) multi-wavelength observations (e.g., radio for jets/outflows and UV or X-rays for winds) further complicates their detection and characterization.

\afhd was re-brightened and identified as a transient on January 4, 2024, by ZTF. It is located in the galaxy LEDA 145386 at a redshift of $z = 0.027$ \citep{Arcavi2024}. Due to its blue continuum, broad Balmer and He II emission lines, and its location at the center of the host galaxy, this re-brightening was classified as a TDE by \cite{Hammerstein2024}. \afhd has also been classified as a Bowen fluorescence flare (BFFs, \citealt{Arcavi2024}). However, its rapid decline--fading by more than 2 magnitudes in the UV and over two orders of magnitude in X-rays within a year--along with its thermal X-ray spectra, contrasts with the behaviors of BFFs (e.g., \citealt{Trakhtenbrot2019,Makrygianni2023,Sniegowska2025}) but is more consistent with that of TDEs (Wang et al. 2025, submitted).
Using the single-epoch method and the M–$\sigma$ relation, the mass of the central black hole was estimated to be $M_{\rm BH} = 10^{6.7\pm0.5}\,M_{\odot}$ (Wang et al. 2025, submitted).
High-cadence radio and X-ray follow-up observations have been scheduled to study the periodic variations observed in both wavelengths, which have been interpreted by Wang et al. (2025, submitted) as evidence for disk–jet co-precession. More importantly, absorption features have been detected in the X-ray observations, making the data set of \afhd particularly well-suited for investigating the above questions.

In this work, we conduct a detailed analysis of the X-ray absorption features observed in \afhd and discuss their implications for wind evolution and the ``missing energy problem".

\section{Data reduction and Method} \label{sec:Methods}
The data reduction procedures of \nicer, \swift, and \xmm/PN follow the methods described in \wangetal. 
The raw data of \xmm/RGS (Obs ID: 0953010101; 0953010201; 0953010301. Hereafter referred to as 101, 201, and 301 for simplicity.) were downloaded from the XMM-Newton Science Archive and processed using XMM-Newton’s Science Analysis Software (version 20.0.0). The source, background spectra, and response files were generated with the {\sc rgsproc} task. We applied the same good time intervals used for background flare removal in \xmm/PN. We stacked the first-order RGS1 and RGS2 with {\sc rgscombine} task to improve the SNR. The energies below 0.3\,keV and above 1\,keV were ignored. The spectra were binned by a factor of 10 in further analysis. The observation log is shown in Tab.~\ref{tab:obs_all}.

\begin{deluxetable*}{lccccccc}[!ht]
\tablecaption{Observation log for \afhd used in this paper.} 
\label{tab:obs_all}
\tablewidth{0pt}
\tablehead{
\colhead{ObsID} & \colhead{Start Time} & \colhead{End Time} & \colhead{Phase$^{a}$} & \colhead{Exposure} & \colhead{Total counts} & \colhead{flux$^{b}$} & \colhead{Outflow} \\[-5pt]
 & (MJD) & (MJD) & (days) & (ks) &  & ($10^{-12}$\,erg\,cm$^{-2}$\,s$^{-1}$) & }
\startdata
\multicolumn{7}{c}{\swift/XRT}\\
\hline
Epoch (1) & 60339.6825 &  60384.2380 & 29.7-74.2 & 31.46 & 1081 & 2.5 & Not detected\\
Epoch (2) & 60482.0168 &  60503.4630 & 172.0-193.4 & 16.91 & 843 & 3.3 & Weak\\
Epoch (3) & 60506.0838 &  60524.6943 & 196.1-214.7 & 12.29 & 566 & 2.7 & Detected\\
Epoch (4) & 60530.1423 &  60611.8776 & 220.1-301.9 & 60.69 & 801 & 0.7 & Not detected\\
\hline
\multicolumn{7}{c}{\nicer}\\
\hline
6705010118 & 60504.3363 &  60504.8634 & 194.3 & 3.2 & 5936 & 3.5 & Detected\\
6705010119 & 60508.0171 &  60508.4761 & 198.0 & 3.2 & 9670 & 4.6 & Detected\\
6705010120 & 60509.0597 &  60509.3858 & 199.1 & 1.3 & 6271 & 9.1 & Detected\\
6705010121 & 60511.0532 &  60511.2603 & 201.1 & 1.7 & 3283 & 4.4 & Detected\\
6705010122$^{c}$ & 60512.9196 & 60512.9365 & 203.0 & 1.2 & \multirow{2}{*}{9793} & \multirow{2}{*}{4.8} & \multirow{2}{*}{Detected}\\
6705010123$^{c}$ & 60512.9842 & 60513.0657 & 203.0 & 2.4 &  & \\
\hline
\multicolumn{7}{c}{\xmm/PN}\\
\hline
0953010201$^{d}$ & 60517.9565 &  60518.2771 & 208.0 & 11.6 & 13018 & 4.0 & Not detected\\
0953010101$^{d}$ & 60521.6477 &  60521.7795 & 211.6 & 10.7 & 14444 & 4.2 & Detected\\
0953010301$^{d}$ & 60525.6214 &  60525.8339 & 215.6 & 16.4 & 1978 & 0.4 & Not detected\\
\enddata
\tablecomments{
$^{a}$: Days since discovery (MJD 60310).\\
$^{b}$: Unabsorbed flux in 0.3--2\,keV.\\
$^{c}$: Given the gap between the two observations is less than 0.1\,days, they were merged in the spectral fitting to improve the statistical significance.\\
$^{d}$: Referred to as \pnb, \pna, and \pnc in this paper.}
\end{deluxetable*}

The spectra were analyzed using {\sc xspec} version 12.12.1 \citep{arnaud1996} and {\sc PYSPEX}\footnote{\href{https://spex-xray.github.io/spex-help/pyspex.html}{https://spex-xray.github.io/spex-help/pyspex.html}} v3.08.01, the python interface of {\sc SPEX} \citep{Kaastra1996}. Spectral fitting for \afhd was performed using C-statistics \citep{Cash1979,Kaastra2017}. The redshift is set at 0.027 \citep{Arcavi2024}. Galactic interstellar absorption was accounted for with a column density fixed at 0.047$\times10^{22}$\,cm$^{-2}$, as determined from the HI4PI survey \cite{HI4PI2016A&A...594A.116H}.

In {\sc xspec}, the abundance tables and photoelectric cross-sections were taken from \cite{Wilms2000} and \cite{Verner1996}, respectively. For \nicer observations, we applied the SCORPEON\footnote{\href{https://heasarc.gsfc.nasa.gov/docs/nicer/analysis_threads/scorpeon-overview/}{https://heasarc.gsfc.nasa.gov/docs/nicer/analysis\_threads/scorpeon-overview/}} model to estimate the background.

In {\sc SPEX}, the SCORPEON background file was generated using the {\sc niscorpspect} task, based on the best-fitted \texttt{nxb} and \texttt{sky} parameters. The spectrum were converted to {\sc SPEX} format using the {\sc ogip2spex} task from the {\sc pyspextools} library. The data were regrouped using the optimal rebinning command \texttt{obin} \citep{Kaastra2016}. We used the proto-solar abundances from \cite{Lodders2009}. All plasma abundances were fixed to solar values.

\begin{figure*}[!ht]
 \centering
   \includegraphics[width=1\linewidth]{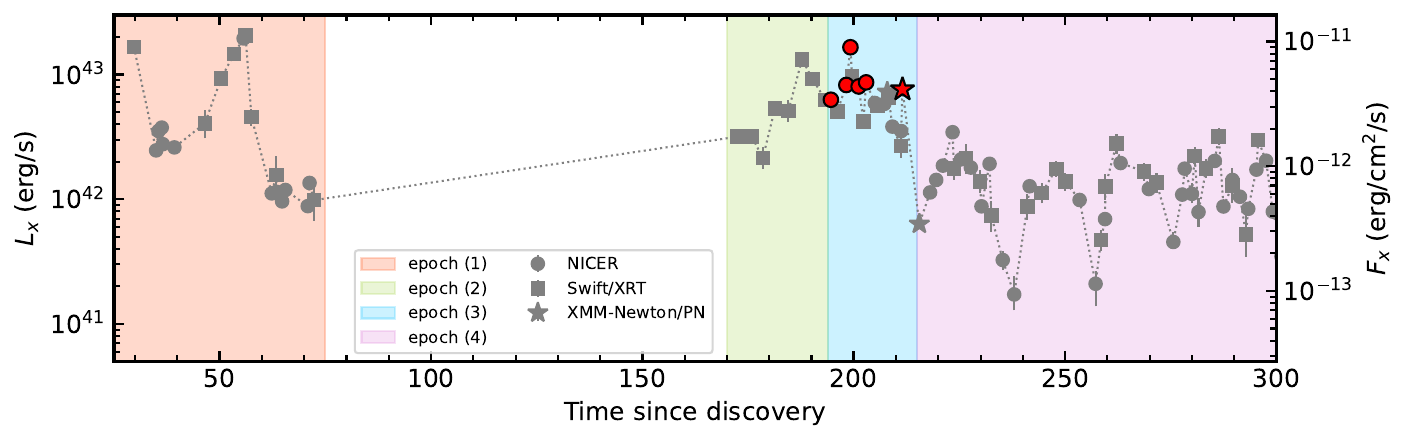}
   \caption{The long-term unabsorbed X-ray (0.3–2\,keV) light curve of \afhd, starting from the discovery date (MJD 60310). Grey circles, squares, and stars represent observations from \nicer, \swift, and \xmm/PN, respectively. Data points in red indicate observations where an absorption feature was detected in the spectrum. The orange, green, blue, and purple shaded regions correspond to the four epochs used to combine the \swift spectra, as described in Sec.~\ref{sec:Result}.}
   \label{fig:lc}
\end{figure*}

To better follow the evolution of the absorption features observed in \afhd (see Sec.~\ref{sec:Result} for details), we split the data into four epochs: (1) 0--74\,days, (2) 172--194\,days, (3) 194--215\,days, and (4) 215--302\,days, corresponding to the colored regions shown in Fig.\ref{fig:lc}. The interval between 74 and 172\,days corresponds to a seasonal gap.

Parameter uncertainties are reported at the 1$\sigma$ confidence level, while upper and lower limits are presented at the 3$\sigma$ confidence level.

\section{Result} \label{sec:Result}

We present an X-ray analysis of \afhd spanning the first 300\,days since its re-brightening. The long-term light curve, provided by \wangetal, is displayed in Figure~\ref{fig:lc}. The continuum was modeled with a multi-color blackbody component \texttt{diskbb} \citep{Mitsuda1984}, with galactic interstellar absorption and redshift accounted for by \texttt{tbabs} and \texttt{zashift}. In the \pnb and \pnc, residuals revealed a hard excess above 1\,keV, which was modeled by an additional \texttt{powerlaw} component. The inclusion of this component significantly improved the fit with \dstat in $54.9$ and $25.8$ for 2 additional degrees of freedom (dof), respectively. We note that the hard excess could alternatively be modeled with a blackbody or disk blackbody component, as these models cannot be clearly distinguished based on current data. For simplicity, we adopt the \texttt{powerlaw} component throughout this paper to describe the hard excess.

\begin{figure}[!h]
 \centering
   \includegraphics[width=1\linewidth]{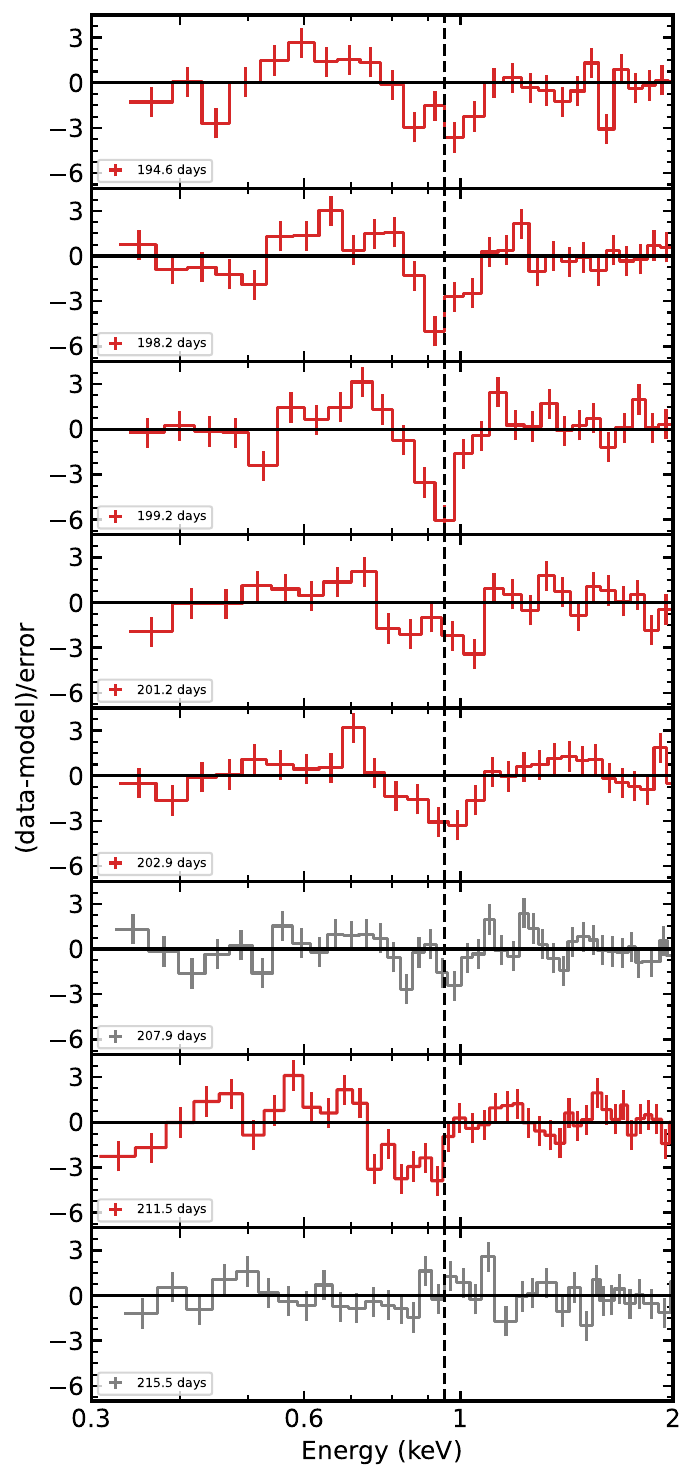}
   \caption{The continuum residuals of the selected \nicer (top five panels) and \xmm/PN (bottom three panels) spectrum. Red curves indicate detections of absorption lines. For comparison, two additional \xmm/PN observations without detected absorption features are shown in gray. The dashed line marks the absorption feature at 0.95\,keV.}
   \label{fig:NICER_XMM_del}
\end{figure}

On top of the continuum, we detected a broad absorption feature near 0.95\,keV in six individual observations from epoch (3), as indicated by the red symbols in Fig.~\ref{fig:lc}. These features can be described by a single Gaussian profile with \dstat$=24$--$66$ for 3 additional dof in five \nicer observations and \pna, as shown in Tab.~\ref{tab:spe_result}. The centroid energy ($E_{\rm line}$) exhibited a prominent shift from 0.95\,keV to 0.78\,keV within roughly 10\,days (see Fig.~\ref{fig:NICER_XMM_del}), potentially indicative of rapid velocity evolution.

To further search for potential absorption features across the dataset, we averaged the residuals of the continuum-subtracted spectra in the 0.8--1.1\,keV range from individual \xmm/PN and \nicer observations (see Appendix Sec.~\ref{sec:abs_feature} for details). Nevertheless, only the six observations highlighted with red symbols in Fig.~\ref{fig:lc} exhibit significant ($> 3\sigma$) deviations from the continuum, consistent with our initial finding. Moreover, given the comparable total counts of individual \nicer observations in epochs (1), (3) and (4) (see Tab.~\ref{tab:nicer_epoch}), we rule out the possibility that the non-detection of absorption features in individual observations from epochs (1) and (4) is due to low SNRs.

\begin{figure*}[!ht]
 \centering
   \includegraphics[width=1\linewidth]{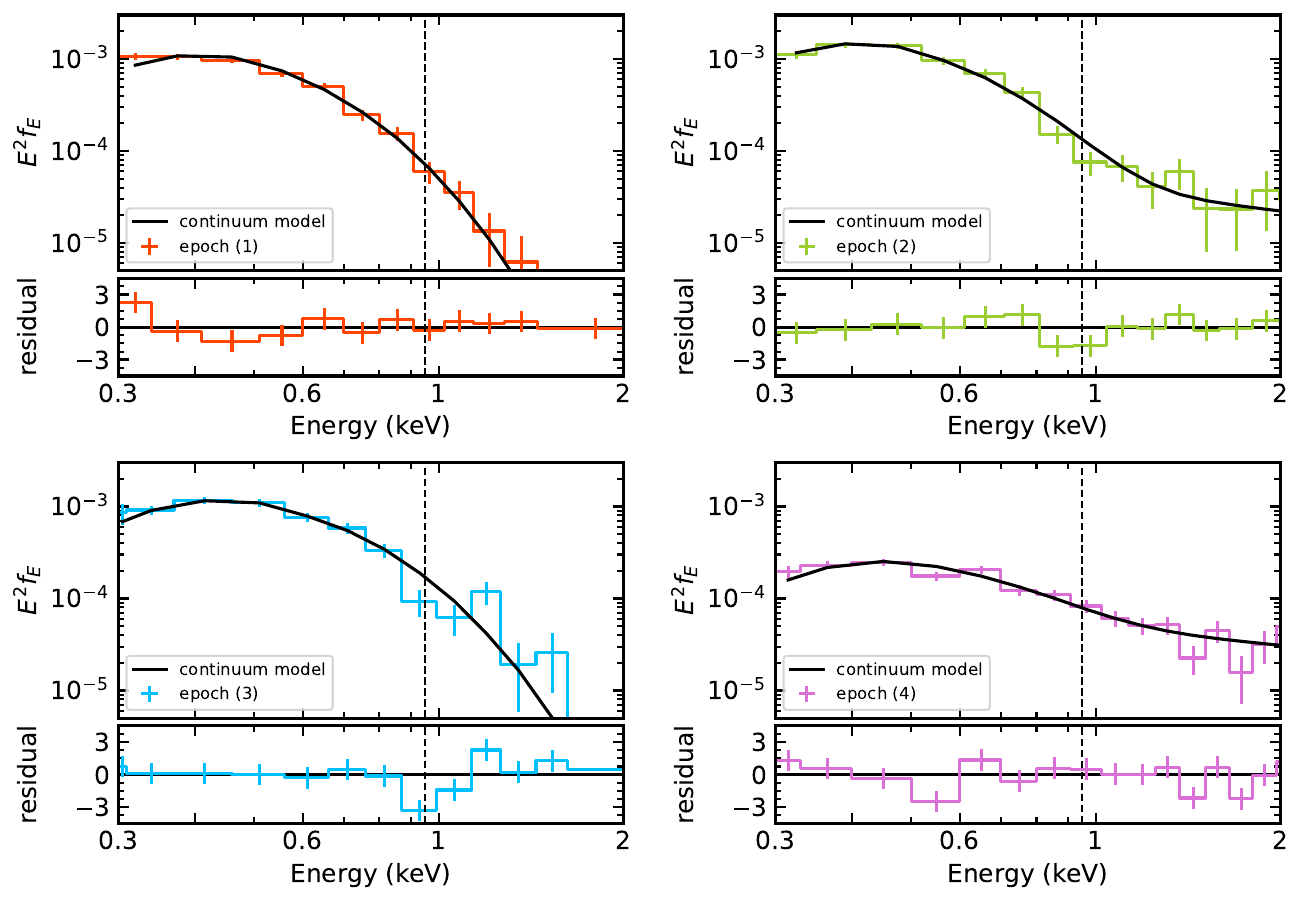}
   \caption{The \swift/XRT continuum fitting spectrum and their corresponding residuals with color code represented same as four epoch in Fig.~\ref{fig:lc}. The dashed lines mark as the absorption feature in 0.95\,keV. Epoch (3) covers the time interval of Fig.~\ref{fig:NICER_XMM_del}.}
   \label{fig:swift_spe}
\end{figure*}

To track the full evolution of the absorption features, we further analyzed the \swift/XRT data, which provides observational coverage in all epochs. The \swift/XRT spectra were combined within each epoch due to the low SNRs of individual observations. In modeling the continuum spectrum, a \texttt{powerlaw} component was required to model the hard excess in epochs (2) and (4). A dip near 0.95\,keV appeared clearly in epochs (2) and (3) but was absent in epochs (1) and (4). The best-fitting continuum is shown in Fig.~\ref{fig:swift_spe}. We note that the estimated background of \nicer may vary from day to day, potentially introducing artificial features when spectra from extended periods, such as tens of days, are merged. Therefore, we did not combine \nicer spectra across epochs in determine the absorption features.

For epochs (2) and (3), we added a Gaussian component to characterize the absorption feature. The line width ($E_{\sigma}$) was unconstrained, with a 1$\sigma$ upper limit of 0.1\,keV, so we fixed it at the median value of 0.05\,keV to improve constraints on other parameters.
To determine upper limits for epoch (1), we fixed $E_{\rm line}$ and $E_{\sigma}$ to the values from epoch (2) to constrain the line strength ($E_{\rm norm}$). For epoch (4), which corresponds to the later stage of line evolution, we fixed $E_{\rm line}$ and $E_{\sigma}$ to the values from \pna. The inclusion of an additional Gaussian component resulted in \dstat = 0, indicating that this component was not statistically required.
We also tested the inclusion of an additional Gaussian component in \pnc, and the results were consistent with epoch (4). In \pnb, which was observed within epoch (3), the additional Gaussian component improved the fit (\dstat = 8.9, with 1 additional dof), but a strong coupling between the Gaussian component and its continuum prevented us from confirming whether the feature was real.
The best-fitting parameters are presented in Tab.~\ref{tab:swift_spe}. These findings indicate that the absorption features appeared intermittently. Moreover, the wind may have emerged as early as 74\,days after discovery (e.g., around day 90 or 150), and became undetectable after day 215.

To achieve a more precise characterization of the observed absorption features, we employed the photoionization model \pion \citep{Miller2015,Mehdipour2016} in {\sc SPEX}. Interstellar medium absorption was accounted for by \texttt{hot} component, and redshift was accounted for by \texttt{reds} component. Additionally, the continuum was modeled using \texttt{dbb}\footnote{\href{https://spex-xray.github.io/spex-help/models/dbb.html}{https://spex-xray.github.io/spex-help/models/dbb.html}} in {\sc SPEX}, which is defined differently from \texttt{diskbb} in {\sc xspec}. When applied to the same spectrum, \texttt{dbb} gives a temperature approximately twice as high as that obtained with \texttt{diskbb}. This discrepancy has also been reported in previous studies \citep[e.g.,][]{Kosec2023,Kosec2025}.
For the \xmm observations, RGS and PN spectra were jointly fitted to improve model parameter constraints. The best-fit parameters are listed in Tab.~\ref{tab:spe_result}. 

\begin{figure}[!t]
 \centering
   \includegraphics[width=1\linewidth]{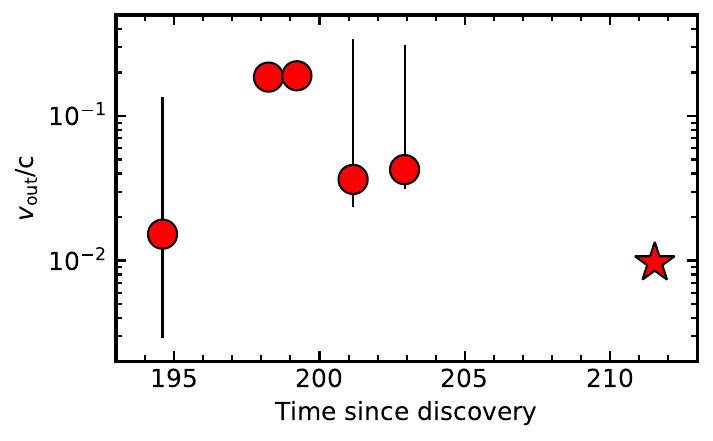}
   \caption{Time evolution of the outflow velocity as determined by the \pion model.}
   \label{fig:vel_evolution}
\end{figure}

Based on the photoionization model, the spectra revealed a maximum blueshift of 0.19c, which gradually decreased to 0.0097c over a timescale of 10\,days (see Fig.~\ref{fig:vel_evolution}). Meanwhile, the column density and ionization parameter of the absorption component evolved in opposite directions (see Fig.~\ref{fig:relation_v_xi_nh}). The broad absorption feature is centered around 0.75\,keV in the rest frame and is contributed by the Iron M-shell Unresolved Transition Array (UTA) and the Oxygen K-edge. To assess the robustness of this result, we compared the fits for \pna when the velocity was allowed to vary freely versus when it was constrained to lie within the velocity range observed in the \nicer spectra (i.e., greater than 10,000\,km/s). The constrained fit yields a significantly worse result, with \dstat$\gtrsim15$, confirming the robustness of the velocity measurement derived from \pna.

\section{Discussion} \label{sec:Discussion}
We searched for absorption features in the X-ray data of the TDE \afhd using observations from Swift/XRT, NICER, and XMM-Newton/PN spanning from approximately 30 to 300\,days after the event. Intermittent absorption features were detected in individual observations taken between 194 and 212\,days post-disruption. 
To mitigate the impact of the limited SNR in individual exposures, we divided the full dataset into four epochs (see Fig.~\ref{fig:lc}) and co-added the XRT data within each epoch. The combined spectra reveal no significant absorption features in epochs (1) and (4), while the absorption strength increases from epoch (2) to epoch (3). Although the exact launch time of the outflow remains uncertain due to a seasonal gap in the data, the appearance of the absorption features is clearly delayed relative to the emergence of the X-ray continuum emission. Moreover, during epoch (3), the absorption features detected in individual observations exhibit a significant decrease in central energies over a span of 10\,days, corresponding to a dramatic velocity drop from $0.19^{+0.02}_{-0.01}$c to $0.0097^{+0.0008}_{-0.0005}$c ($\sim2900$\,km\,s$^{-1}$). 
In the following, we compare our results with absorption features detected in other TDEs, discuss the possible driving mechanisms, and explore the broader implications of our findings.

\subsection{Comparison with absorption features detected in other TDEs} \label{sec:geometry}

Currently, blue-shifted X-ray absorption features have been observed in only a few TDEs, all of which exhibit ultra-fast outflow (UFO) components with velocities of $\sim$0.1–0.3c. Notably, ASASSN-14li \citep{Kara2018} and ASASSN-20qc \citep{Pasham2024} detected UFOs during the early phase, within $\lesssim$30\,days of discovery. In contrast, UFO components have also been identified at later times, approximately 150, 230, and 300\,days post-discovery, in 3XMM J152130.7+074916 \citep{Lin2015ApJ...811...43L}, AT2020ksf \citep{Wevers2023}, and AT2021ehb \citep{Xiang2024}, respectively. Moreover, in AT2020ksf, absorption features remained detectable even 770\,days after discovery \citep{Wevers2023}, although the significant observational gap makes it uncertain whether the outflows were persistently present throughout this period.
Due to the lack of early-time observations or high SNR spectra, it remains unclear whether the disk winds were launched at earlier stages. In contrast, for \afhd, it is evident that the UFO was launched no earlier than 74\,days post-discovery, strengthened between days 172 and 194, and disappeared after day 215, marking the first instance in which the full evolutionary process of outflows has been observed in a TDE.

Regarding the UV absorption features detected in TDEs, most of them were observed within 60\,days of discovery \citep{Chornock2014,Cenko2016,Brown2018,Blagorodnova2019,Hung2019,Hung2021,Payne2023}, with the exception of PS16dtm \citep{Blanchard2017}, for which the UV observation were conducted 191\,days post-discovery. Moreover, the UV outflows tend to be statistically slower than the X-ray ones, with typical velocities of $\lesssim 0.05$c.
In the case of \afhd, \textit{Hubble Space Telescope} (\hst) observations were conducted on days 56 and 258 post-discovery, but no UV absorption features were detected (private communication with the PI of the \hst observations, Ning Jiang).
Since UV spectroscopic resources are extremely limited compared to other bands, UV coverage remains sparse, which in turn limits our understanding of the evolution of UV outflows.

Additionally, the outflows in \afhd exhibited a dramatic deceleration from 0.19c to 0.0097c within approximately 10\,days. 
Outflow deceleration, from 15,000\,km\,s$^{-1}$ to 10,000\,km\,s$^{-1}$, was observed in the UV band of the TDE AT2019qiz over the course of 52\,days \citep{Hung2021}, and was attributed to a decline in the mass fallback rate. However, the outflow deceleration in \afhd occurred over a much shorter timescale, while the luminosity remained at the same level, suggesting a different origin.
Possible explanations for the substantial slowdown in \afhd include: (1) a collision between the UFO and a slower outflowing component \citep{Yang2023}; (2) a change in the origin of the outflow, with the slower component being launched from a different region of the disk than the initial UFO. 
The observed increase in the ionization parameter during the deceleration disfavours the collision scenario. Therefore, the deceleration is more likely attributed to the intrinsic properties of the wind, possibly governed by the underlying driving mechanisms, which we discuss in detail in Section~\ref{sec:origin}.

\begin{figure*}[!ht]
 \centering
   \includegraphics[width=1\linewidth]{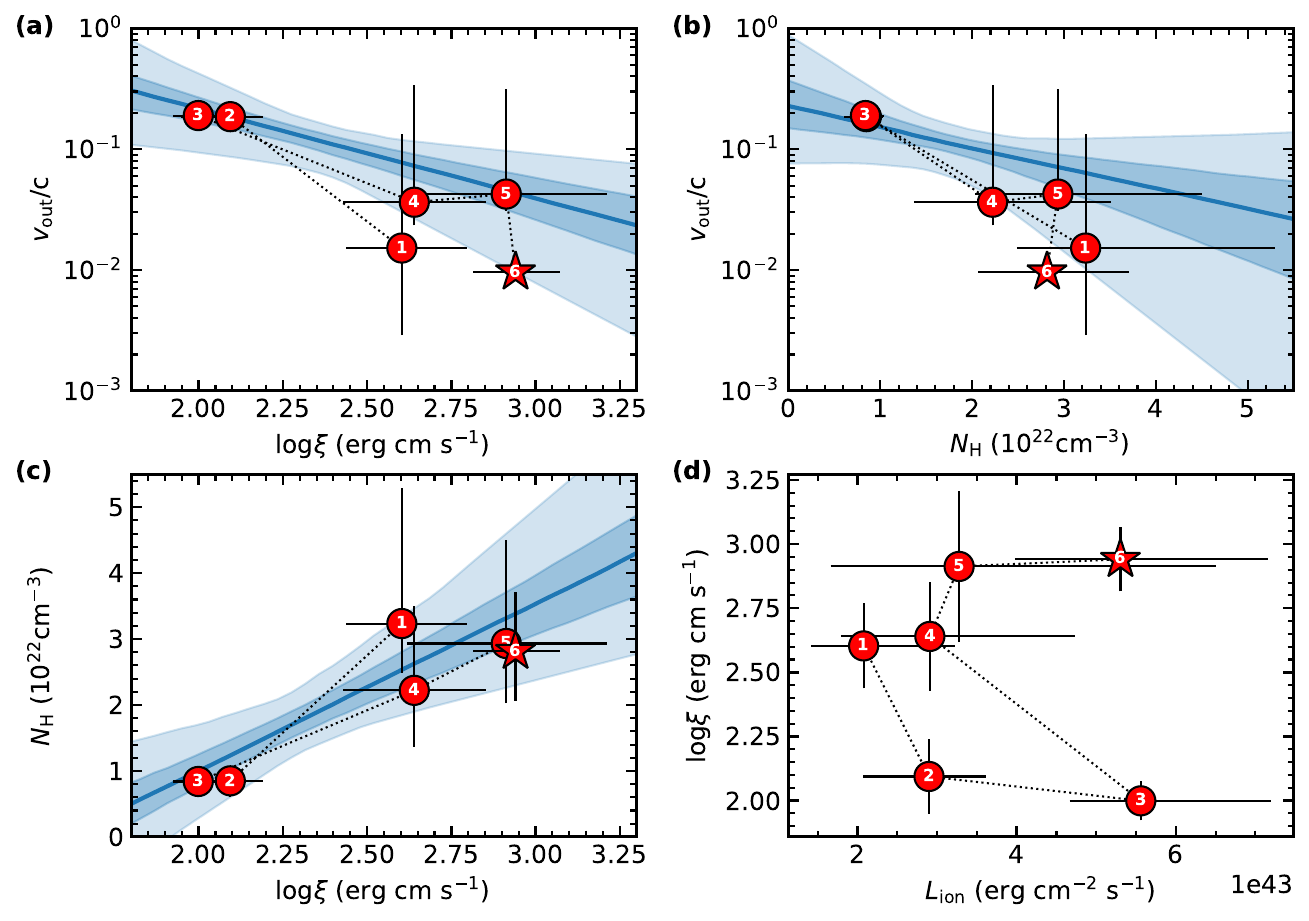}
   \caption{Best-fitting model parameters for the \texttt{hot}$\times$\texttt{reds}$\times$\texttt{pion}$\times$\texttt{dbb} model applied to \nicer and \xmm observations. Panel (a): Outflow velocity vs. ionization parameter. Panel (b): Outflow velocity vs. absorber column density. Panel (c): Absorber column density vs. ionization parameter. Panel (d): Ionization parameter vs. ionizing luminosity (0.0136--13.6\,keV). The numbers on the data points indicate their time sequence in the evolution. The blue lines represent the best-fitting linear relationships, while the dark and light blue regions indicate the 1$\sigma$ and 3$\sigma$ confidence intervals, respectively.}
   \label{fig:relation_v_xi_nh}
\end{figure*}

\subsection{Driving mechanisms of the outflow} \label{sec:origin}

For SMBHs with masses below $3 \times 10^7\,M_{\odot}$ that disrupt solar-like stars, the resulting events typically exhibit super-Eddington accretion during the early stages \citep{Strubbe2009,Lodato2011,Metzger2016,Gezari2021}. In the super-Eddington accretion regime of TDEs, the thermal luminosity is expected to remain roughly constant as the accretion rate declines \citep{Krolik2012}. Once the flow transitions to a sub-Eddington, radiatively efficient state, the luminosity begins to decrease.
This predicted constant-luminosity phase has been observed in TDEs such as Swift J2058.4+0516 \citep{Cenko2012,Krolik2012}, 3XMM J215022.4-055108 \citep{Lin2018}, and EP240222a \citep{Jin2025}. Therefore, the two plateaus observed within the first year following the re-brightening of \afhd are likely indicative of sustained super-Eddington or Eddington-limited accretion. 

In super-Eddington accreting phase, strong radiation pressure is expected to drive the formation of optically thick, wide-angle winds \citep[see e.g.,][]{Ohsuga2005,Strubbe2009,Jiang2014,McKinney2014,Dai2018}. Radiation pressure-driven winds are expected to be mildly ionized and to exhibit a positive correlation among outflow velocity, ionization parameter, and luminosity. Such correlations have been observed in several ultraluminous X-ray sources \citep[e.g.,][]{Pinto2020MNRAS.491.5702P} and active galactic nuclei \citep[e.g.,][]{Matzeu2017,Pinto2018,Xu2024}.
While in the case of \afhd, our observations reveal an opposite trend between outflow velocity and ionization parameter, as well as no clear correlation between the ionization parameter and ionizing luminosity (see Fig.~\ref{fig:relation_v_xi_nh}). Moreover, the delayed onset and subsequent strengthening of the outflows are also inconsistent with the radiation pressure-driven wind scenario in TDEs, where outflows are expected to weaken as the mass accretion rate declines. Other mechanisms are therefore needed to account for the observed absorption features.

Thermal winds are expected to arise when X-ray heating raises the gas temperature such that the sound speed exceeds the local escape velocity \citep{Begelman1983}. This condition is typically met when the wind-launching radius exceeds the Compton radius, defined as $R_{\rm C} = (10^{10}/T_{\rm C,8})\times(M_{\rm BH}/M_\odot)$\,cm, where $T_{\rm C,8}$ is the Compton temperature in units of $10^{8}$\,K \citep{Begelman1983,Wang2021ApJ...906...11W}. Using the disk temperature derived from continuum fitting, we estimate $R_{\rm C} \sim 5\times10^{18}$\,cm.
The wind-launching radius can be approximated by $R_{\rm launch} = \sqrt{L_{\rm ion}/n_{\rm gas}\xi_{\rm gas}}$, where $L_{\rm ion}$ is the ionizing luminosity in the 0.0136--13.6\,keV band and $n_{\rm gas}$ is hydrogen number density. Based on hydrodynamical simulations \citep[e.g.][]{Bu2022MNRAS, Thomsen2022}, we adopt a conservative range of $n_{\rm gas} = 10^{10}$ to $10^{15}$\,cm$^{-3}$, resulting in $R_{\rm launch} = 10^{13}$ to $10^{16}$\,cm. These values are at least two orders of magnitude smaller than the Compton radius, suggesting that the observed disk wind is unlikely to be thermally driven.

Magnetically driven winds primarily depend on the magnetic field, such that the outflow velocity could be independent of the ionization state \citep{Wang2022MNRAS.513.5818W}. 
This could in principle explain the observed deceleration in \afhd.
Characteristic asymmetries in absorption line profiles may serve as diagnostic signatures of such winds \citep{Fukumura2022,Chakravorty2023, Datta2024}. However, the current spectral resolution of available observations limits our ability to robustly confirm this scenario.

\subsection{Plausible explanations for the delayed onset outflow}

We propose three plausible explanations for the delayed onset of the observed outflow in \afhd:
(a) The outflow may have been highly ionized in the early phase, making electron scattering the dominant process over absorption. As the ionization parameter decreased over time, absorption lines became more prominent and detectable.
(b) The outflow could have initially been confined within a narrow opening angle; as the angle widened, the wind began to intersect our line of sight.
(c) Heavy elements, such as iron, may have gradually built up in the accretion disk as the core of the disrupted star was eventually accreted onto the central black hole.

In scenario (a), if outflows were launched during the early epoch via the same driving mechanism as in later epochs, the corresponding launching radius would likely remain within a similar range. Given that the luminosity stayed in a plateau phase during the first 215\,days in \afhd, the ionization parameter would primarily depend on the gas density ($\xi \propto n_{\rm gas}^{-1}$). Since the mass fallback rate in TDEs generally decreases over time \citep{Rees1988Nat, Phinney1989}, the gas density is also expected to decline, leading to an increase in the ionization parameter rather than a decrease. Therefore, we do not favor this scenario. Regarding scenario (b), it conflicts with unified TDE models, which suggest that outflows are primarily driven by radiation pressure during the early stages and are expected to exhibit large opening angles \citep{Ohsuga2005, Jiang2014, Dai2018}. However, the observed properties of the outflows in \afhd are inconsistent with a radiation pressure-driven mechanism.
We thus could not rule out this possibility. 
For scenario (c), since the star is disrupted from the outside in, the core—richer in heavy elements—would be disrupted and accreted onto the black hole later than the outer layers. This delayed accretion of metal-rich material could contribute to the emergence of absorption features. However, during disk formation, the stellar debris undergoes strong self-interactions and collisions \citep{Bonnerot2017, Andalman2022}, resulting in a chaotic redistribution of the elements. This makes it difficult to assign a precise physical meaning to the observed onset time of the outflow. For instance, in the case of TDE ASASSN-14li \citep{Kara2018}, the outflow was detected just $\sim7$\,days after the disruption—significantly earlier than in \afhd. 
Overall, the delayed onset of the outflow in \afhd may be attributed to an increasing opening angle and/or gradual metal enrichment in the accretion disk.

\subsection{Mass loss rate} \label{sec:massloss}

To estimate the mass loss rate from the outflow and address the ``missing energy problem" in TDEs, we adopted the following relation from \cite{Nardini2015},
\begin{align}
\dot M_{\rm wind} = \Omega m_{\rm p} N_{\rm H} v_{\rm out} R_{\rm launch}
\end{align}
We assumed a solid angle of $\Omega = 2\pi$ and the maximum outflow velocity of 0.19c. Considering that $R_{\rm launch}$ spans a broad range from 10 to 1000\,$R_{\rm g}$, where $R_{\rm g} = GM_{\rm BH}/c^2$, we estimated a mass outflow rate of 4--400$\times10^{21}$\,g\,s$^{-1}$. Even if the outflow persists for 100\,days, the total mass loss remains below $0.002\,M_{\odot}$.

Significant radio emission was also detected from \afhd, with fluxes ranging between $\sim2.5 \times 10^{37}$\,erg\,s$^{-1}$ and $\sim3.1 \times 10^{38}$\,erg\,s$^{-1}$ over the studied period (Wang et al. 2025, submitted). Similar to the X-rays, the radio emission also exhibited high-amplitude, short-term variability, with the flux varying by up to a factor of four on timescales of 10--30\,days.
Such rapid variations are difficult to explain within the framework of isotropic winds. Moreover, a strong correlation between the radio and X-ray light curves was detected, suggesting that the radio emission most likely originates from a jet, as discussed in \wangetal.
Therefore, we also estimated the mass loss rate associated with the jet using a simple model in which electrons cross an emitting region with a cross-sectional area $A = \pi R_{\rm blob}^2$ at the jet velocity $v_{\rm jet}$. Assuming charge neutrality, the number of ejected protons equals the number of electrons, leading to the following expression for the mass loss rate:
\begin{align}
    \dot M_{\rm jet} = \rho vA = m_{\rm p}n_{\rm e} v_{\rm jet} \pi R_{\rm blob}^2
\end{align}
where $R_{\rm blob}$ and $n_{\rm e}$ are the radius and electron number density of the jet-emitting region, respectively. Under the equipartition assumption discussed in \wangetal, we adopt $v_{\rm jet}=0.6$c, $R_{\rm blob} \sim 1.1-3.6\times10^{16}$\,cm and $n_{\rm e} \sim 1.2-18.1$\,cm$^{-3}$, which yields a mass loss rate of approximately $\sim 2\times10^{20}$\,g\,s$^{-1}$. Although the derived mass loss rate is at least an order of magnitude lower than $\dot{M}_{\rm wind}$, the jet activity persists up to the time of writing, and its termination time remains uncertain. 

Thus, combining the contributions from both the disk wind and the jet, the total mass outflow constitutes only a minor fraction of the mass of the disrupted solar-like star. However, this estimate likely represents a lower limit on the actual mass loss, as the detectability of outflows strongly depends on the viewing angle. It is possible that a significant portion of the outflows is launched outside our line of sight. Therefore, even with a dataset as densely sampled as that of \afhd, it remains unlikely that observations alone can confirm the role of outflows in resolving the ``missing energy problem".

\section{CONCLUSIONS} \label{sec:CONCLUSIONS}

\afhd has exhibited intermittent absorption features, specifically the iron UTA and oxygen K-edge, in the X-ray spectra obtained with \swift, \nicer, and \xmm.

The outflow was detected 142\,days later than the X-ray continuum and showed subsequent strengthening after its discovery. Morever, the outflow underwent a rapid velocity decline from $\sim$0.19c to $\sim$0.0097c over the course of approximately 10\,days, with its velocity found to be anti-correlated with the column density. These observational features, along with our calculations, rule out radiation pressure and thermal mechanisms as the dominant driving forces, leaving magnetic processes as the most viable explanation. High-resolution X-ray observatories, such as XRISM \citep{Tashiro2022}, and future missions like HUBS \citep{HUBS2020} and Athena \citep{Barret2023}, will be crucial in constraining the physical nature of these winds in bright TDEs.

We suggest that the delayed onset of the outflow may be linked to an increase in the wind’s opening angle and/or to metal enrichment (e.g., iron and oxygen) during the disk formation phase.

Regarding the ``missing energy problem", our results show that the detected outflows, including both the jet and disk wind, account for only a small fraction of the total mass of the disrupted star. Observations are inevitably limited by viewing angles, and as such, may not be able to fully assess the role of outflows in resolving the ``missing energy problem". 

\begin{acknowledgments}

We thank the anonymous referee for the constructive comments, which have helped improve the manuscript. 
We thank Erlin Qiao, Yiyang Lin, and Yongxin Wu for the useful discussion. We thank the \nicer Helpdesk for their guidance on using the SCORPEON model. This research was supported by the National Natural Science Foundation of China (NSFC) under grant number 12588202; by the New Cornerstone Science Foundation through the New Cornerstone Investigator Program and the XPLORER PRIZE; and by the Strategic Priority Program of the Chinese Academy of Sciences under grant numbers XDB41000000 and XDB0550203. D. Bu is supported by the Natural Science Foundation of China (grants 12173065, 12133008, 12192220, 12192223).

\end{acknowledgments}

\bibliography{main}{}

\begin{thebibliography}{}
\expandafter\ifx\csname natexlab\endcsname\relax\def\natexlab#1{#1}\fi
\providecommand{\url}[1]{\href{#1}{#1}}
\providecommand{\dodoi}[1]{doi:~\href{http://doi.org/#1}{\nolinkurl{#1}}}
\providecommand{\doeprint}[1]{\href{http://ascl.net/#1}{\nolinkurl{http://ascl.net/#1}}}
\providecommand{\doarXiv}[1]{\href{https://arxiv.org/abs/#1}{\nolinkurl{https://arxiv.org/abs/#1}}}

\bibitem[{{Andalman} {et~al.}(2022){Andalman}, {Liska}, {Tchekhovskoy}, {Coughlin}, \& {Stone}}]{Andalman2022}
{Andalman}, Z.~L., {Liska}, M. T.~P., {Tchekhovskoy}, A., {Coughlin}, E.~R., \& {Stone}, N. 2022, \mnras, 510, 1627, \dodoi{10.1093/mnras/stab3444}

\bibitem[{{Arcavi} {et~al.}(2024){Arcavi}, {Faris}, {Newsome}, {Sniegowska}, \& {Trakhtenbrot}}]{Arcavi2024}
{Arcavi}, I., {Faris}, S., {Newsome}, M., {Sniegowska}, M., \& {Trakhtenbrot}, B. 2024, Transient Name Server Classification Report, 2024-466, 1

\bibitem[{{Arnaud}(1996)}]{arnaud1996}
{Arnaud}, K.~A. 1996, in Astronomical Society of the Pacific Conference Series, Vol. 101, Astronomical Data Analysis Software and Systems V, ed. G.~H. {Jacoby} \& J.~{Barnes}, 17

\bibitem[{{Barret} {et~al.}(2023){Barret}, {Albouys}, {Herder}, {Piro}, {Cappi}, {Huovelin}, {Kelley}, {Mas-Hesse}, {Paltani}, {Rauw}, {Rozanska}, {Svoboda}, {Wilms}, {Yamasaki}, {Audard}, {Bandler}, {Barbera}, {Barcons}, {Bozzo}, {Ceballos}, {Charles}, {Costantini}, {Dauser}, {Decourchelle}, {Duband}, {Duval}, {Fiore}, {Gatti}, {Goldwurm}, {Hartog}, {Jackson}, {Jonker}, {Kilbourne}, {Korpela}, {Macculi}, {Mendez}, {Mitsuda}, {Molendi}, {Pajot}, {Pointecouteau}, {Porter}, {Pratt}, {Pr{\^e}le}, {Ravera}, {Sato}, {Schaye}, {Shinozaki}, {Skup}, {Soucek}, {Thibert}, {Vink}, {Webb}, {Chaoul}, {Raulin}, {Simionescu}, {Torrejon}, {Acero}, {Branduardi-Raymont}, {Ettori}, {Finoguenov}, {Grosso}, {Kaastra}, {Mazzotta}, {Miller}, {Miniutti}, {Nicastro}, {Sciortino}, {Yamaguchi}, {Beaumont}, {Cucchetti}, {D'Andrea}, {Eckart}, {Ferrando}, {Kammoun}, {Lotti}, {Mesnager}, {Natalucci}, {Peille}, {de Plaa}, {Ardellier}, {Argan}, {Bellouard}, {Carron}, {Cavazzuti}, {Fiorini}, {Khosropanah}, {Martin}, {Perry}, {Pinsard},
  {Pradines}, {Rigano}, {Roelfsema}, {Schwander}, {Torrioli}, {Ullom}, {Vera}, {Villegas}, {Zuchniak}, {Brachet}, {Cicero}, {Doriese}, {Durkin}, {Fioretti}, {Geoffray}, {Jacques}, {Kirsch}, {Smith}, {Adams}, {Gloaguen}, {Hoogeveen}, {van der Hulst}, {Kiviranta}, {van der Kuur}, {Ledot}, {van Leeuwen}, {van Loon}, {Lyautey}, {Parot}, {Sakai}, {van Weers}, {Abdoelkariem}, {Adam}, {Adami}, {Aicardi}, {Akamatsu}, {Alonso}, {Amato}, {Andr{\'e}}, {Angelinelli}, {Anon-Cancela}, {Anvar}, {Atienza}, {Attard}, {Auricchio}, {Balado}, {Bancel}, {Barusso}, {Bascu{\~n}an}, {Bernard}, {Berrocal}, {Blin}, {Bonino}, {Bonnet}, {Bonny}, {Boorman}, {Boreux}, {Bounab}, {Boutelier}, {Boyce}, {Brienza}, {Bruijn}, {Bulgarelli}, {Calarco}, {Callanan}, {Campello}, {Camus}, {Canourgues}, {Capobianco}, {Cardiel}, {Castellani}, {Cheatom}, {Chervenak}, {Chiarello}, {Clerc}, {Clerc}, {Cobo}, {Coeur-Joly}, {Coleiro}, {Colonges}, {Corcione}, {Coriat}, {Coynel}, {Cuttaia}, {D'Ai}, {D'anca}, {Dadina}, {Daniel}, {Dauner}, {DeNigris},
  {Dercksen}, {DiPirro}, {Doumayrou}, {Dubbeldam}, {Dupieux}, {Dupourqu{\'e}}, {Durand}, {Eckert}, {Eiriz}, {Ercolani}, {Etcheverry}, {Finkbeiner}, {Fiocchi}, {Fossecave}, {Franssen}, {Frericks}, {Gabici}, {Gant}, {Gao}, {Gastaldello}, \& {Genolet}}]{Barret2023}
{Barret}, D., {Albouys}, V., {Herder}, J.-W.~d., {et~al.} 2023, Experimental Astronomy, 55, 373, \dodoi{10.1007/s10686-022-09880-7}

\bibitem[{{Begelman} {et~al.}(1983){Begelman}, {McKee}, \& {Shields}}]{Begelman1983}
{Begelman}, M.~C., {McKee}, C.~F., \& {Shields}, G.~A. 1983, \apj, 271, 70, \dodoi{10.1086/161178}

\bibitem[{{Blagorodnova} {et~al.}(2019){Blagorodnova}, {Cenko}, {Kulkarni}, {Arcavi}, {Bloom}, {Duggan}, {Filippenko}, {Fremling}, {Horesh}, {Hosseinzadeh}, {Karamehmetoglu}, {Levan}, {Masci}, {Nugent}, {Pasham}, {Veilleux}, {Walters}, {Yan}, \& {Zheng}}]{Blagorodnova2019}
{Blagorodnova}, N., {Cenko}, S.~B., {Kulkarni}, S.~R., {et~al.} 2019, \apj, 873, 92, \dodoi{10.3847/1538-4357/ab04b0}

\bibitem[{{Blanchard} {et~al.}(2017){Blanchard}, {Nicholl}, {Berger}, {Guillochon}, {Margutti}, {Chornock}, {Alexander}, {Leja}, \& {Drout}}]{Blanchard2017}
{Blanchard}, P.~K., {Nicholl}, M., {Berger}, E., {et~al.} 2017, \apj, 843, 106, \dodoi{10.3847/1538-4357/aa77f7}

\bibitem[{{Bonnerot} {et~al.}(2017){Bonnerot}, {Rossi}, \& {Lodato}}]{Bonnerot2017}
{Bonnerot}, C., {Rossi}, E.~M., \& {Lodato}, G. 2017, \mnras, 464, 2816, \dodoi{10.1093/mnras/stw2547}

\bibitem[{{Brown} {et~al.}(2018){Brown}, {Kochanek}, {Holoien}, {Stanek}, {Auchettl}, {Shappee}, {Prieto}, {Morrell}, {Falco}, {Strader}, {Chomiuk}, {Post}, {Villanueva}, {Mathur}, {Dong}, {Chen}, \& {Bose}}]{Brown2018}
{Brown}, J.~S., {Kochanek}, C.~S., {Holoien}, T.~W.~S., {et~al.} 2018, \mnras, 473, 1130, \dodoi{10.1093/mnras/stx2372}

\bibitem[{{Bu} {et~al.}(2022){Bu}, {Qiao}, {Yang}, {Liu}, {Chen}, \& {Wu}}]{Bu2022MNRAS}
{Bu}, D.-F., {Qiao}, E., {Yang}, X.-H., {et~al.} 2022, \mnras, 516, 2833, \dodoi{10.1093/mnras/stac2399}

\bibitem[{{Cash}(1979)}]{Cash1979}
{Cash}, W. 1979, \apj, 228, 939, \dodoi{10.1086/156922}

\bibitem[{{Cenko} {et~al.}(2012){Cenko}, {Krimm}, {Horesh}, {Rau}, {Frail}, {Kennea}, {Levan}, {Holland}, {Butler}, {Quimby}, {Bloom}, {Filippenko}, {Gal-Yam}, {Greiner}, {Kulkarni}, {Ofek}, {Olivares E.}, {Schady}, {Silverman}, {Tanvir}, \& {Xu}}]{Cenko2012}
{Cenko}, S.~B., {Krimm}, H.~A., {Horesh}, A., {et~al.} 2012, \apj, 753, 77, \dodoi{10.1088/0004-637X/753/1/77}

\bibitem[{{Cenko} {et~al.}(2016){Cenko}, {Cucchiara}, {Roth}, {Veilleux}, {Prochaska}, {Yan}, {Guillochon}, {Maksym}, {Arcavi}, {Butler}, {Filippenko}, {Fruchter}, {Gezari}, {Kasen}, {Levan}, {Miller}, {Pasham}, {Ramirez-Ruiz}, {Strubbe}, {Tanvir}, \& {Tombesi}}]{Cenko2016}
{Cenko}, S.~B., {Cucchiara}, A., {Roth}, N., {et~al.} 2016, \apjl, 818, L32, \dodoi{10.3847/2041-8205/818/2/L32}

\bibitem[{{Chakravorty} {et~al.}(2023){Chakravorty}, {Petrucci}, {Datta}, {Ferreira}, {Wilms}, {Jacquemin-Ide}, {Clavel}, {Marcel}, {Rodriguez}, {Malzac}, {Belmont}, {Corbel}, {Coriat}, {Henri}, \& {Parra}}]{Chakravorty2023}
{Chakravorty}, S., {Petrucci}, P.-O., {Datta}, S.~R., {et~al.} 2023, \mnras, 518, 1335, \dodoi{10.1093/mnras/stac2835}

\bibitem[{{Chornock} {et~al.}(2014){Chornock}, {Berger}, {Gezari}, {Zauderer}, {Rest}, {Chomiuk}, {Kamble}, {Soderberg}, {Czekala}, {Dittmann}, {Drout}, {Foley}, {Fong}, {Huber}, {Kirshner}, {Lawrence}, {Lunnan}, {Marion}, {Narayan}, {Riess}, {Roth}, {Sanders}, {Scolnic}, {Smartt}, {Smith}, {Stubbs}, {Tonry}, {Burgett}, {Chambers}, {Flewelling}, {Hodapp}, {Kaiser}, {Magnier}, {Martin}, {Neill}, {Price}, \& {Wainscoat}}]{Chornock2014}
{Chornock}, R., {Berger}, E., {Gezari}, S., {et~al.} 2014, \apj, 780, 44, \dodoi{10.1088/0004-637X/780/1/44}

\bibitem[{{Cui} {et~al.}(2020){Cui}, {Bregman}, {Bruijn}, {Chen}, {Chen}, {Cui}, {Fang}, {Gao}, {Gao}, {Gao}, {Gottardi}, {Gu}, {Guo}, {Guo}, {He}, {He}, {den Herder}, {Huang}, {Li}, {Li}, {Li}, {Li}, {Li}, {Li}, {Liang}, {Liang}, {Liang}, {Liu}, {Liu}, {Liu}, {Jaeckel}, {Ji}, {Ji}, {Jin}, {Kang}, {Ma}, {McCammon}, {Mo}, {Nagayoshi}, {Nelms}, {Qi}, {Quan}, {Ridder}, {Shen}, {Simionescu}, {Taralli}, {Wang}, {Wang}, {Wang}, {Wang}, {Wang}, {Wang}, {Wang}, {Wang}, {Wang}, {Wang}, {Wang}, {Wang}, {Wang}, {Wang}, {Wen}, {de Wit}, {Wu}, {Xu}, {Xu}, {Xu}, {Xu}, {Xu}, {Xue}, {Yi}, {Yu}, {Yang}, {Yuan}, {Zhang}, {Zhang}, {Zhang}, {Zhong}, {Zhou}, \& {Zhu}}]{HUBS2020}
{Cui}, W., {Bregman}, J.~N., {Bruijn}, M.~P., {et~al.} 2020, in Society of Photo-Optical Instrumentation Engineers (SPIE) Conference Series, Vol. 11444, Space Telescopes and Instrumentation 2020: Ultraviolet to Gamma Ray, ed. J.-W.~A. {den Herder}, S.~{Nikzad}, \& K.~{Nakazawa}, 114442S, \dodoi{10.1117/12.2560871}

\bibitem[{{Curd} \& {Narayan}(2019)}]{Curd2019}
{Curd}, B., \& {Narayan}, R. 2019, \mnras, 483, 565, \dodoi{10.1093/mnras/sty3134}

\bibitem[{{Dai} {et~al.}(2018){Dai}, {McKinney}, {Roth}, {Ramirez-Ruiz}, \& {Miller}}]{Dai2018}
{Dai}, L., {McKinney}, J.~C., {Roth}, N., {Ramirez-Ruiz}, E., \& {Miller}, M.~C. 2018, \apjl, 859, L20, \dodoi{10.3847/2041-8213/aab429}

\bibitem[{{Datta} {et~al.}(2024){Datta}, {Chakravorty}, {Ferreira}, {Petrucci}, {Kallman}, {Jacquemin-Ide}, {Zimniak}, {Wilms}, {Bianchi}, {Parra}, \& {Clavel}}]{Datta2024}
{Datta}, S.~R., {Chakravorty}, S., {Ferreira}, J., {et~al.} 2024, \aap, 687, A2, \dodoi{10.1051/0004-6361/202349129}

\bibitem[{{Fukumura} {et~al.}(2022){Fukumura}, {Dadina}, {Matzeu}, {Tombesi}, {Shrader}, \& {Kazanas}}]{Fukumura2022}
{Fukumura}, K., {Dadina}, M., {Matzeu}, G., {et~al.} 2022, \apj, 940, 6, \dodoi{10.3847/1538-4357/ac9388}

\bibitem[{{Gezari}(2021)}]{Gezari2021}
{Gezari}, S. 2021, \araa, 59, 21, \dodoi{10.1146/annurev-astro-111720-030029}

\bibitem[{{Guillochon} \& {Ramirez-Ruiz}(2013)}]{Guillochon2013}
{Guillochon}, J., \& {Ramirez-Ruiz}, E. 2013, \apj, 767, 25, \dodoi{10.1088/0004-637X/767/1/25}

\bibitem[{{Hammerstein} {et~al.}(2024){Hammerstein}, {Chornock}, {Gezari}, \& {Yao}}]{Hammerstein2024}
{Hammerstein}, E., {Chornock}, R., {Gezari}, S., \& {Yao}, Y. 2024, Transient Name Server Classification Report, 2024-336, 1

\bibitem[{{HI4PI Collaboration} {et~al.}(2016){HI4PI Collaboration}, {Ben Bekhti}, {Fl{\"o}er}, {Keller}, {Kerp}, {Lenz}, {Winkel}, {Bailin}, {Calabretta}, {Dedes}, {Ford}, {Gibson}, {Haud}, {Janowiecki}, {Kalberla}, {Lockman}, {McClure-Griffiths}, {Murphy}, {Nakanishi}, {Pisano}, \& {Staveley-Smith}}]{HI4PI2016A&A...594A.116H}
{HI4PI Collaboration}, {Ben Bekhti}, N., {Fl{\"o}er}, L., {et~al.} 2016, \aap, 594, A116, \dodoi{10.1051/0004-6361/201629178}

\bibitem[{{Hills}(1975)}]{Hills1975}
{Hills}, J.~G. 1975, \nat, 254, 295, \dodoi{10.1038/254295a0}

\bibitem[{{Hung} {et~al.}(2019){Hung}, {Cenko}, {Roth}, {Gezari}, {Veilleux}, {van Velzen}, {Gaskell}, {Foley}, {Blagorodnova}, {Yan}, {Graham}, {Brown}, {Siebert}, {Frederick}, {Ward}, {Gatkine}, {Gal-Yam}, {Yang}, {Schulze}, {Dimitriadis}, {Kupfer}, {Shupe}, {Rusholme}, {Masci}, {Riddle}, {Soumagnac}, {van Roestel}, \& {Dekany}}]{Hung2019}
{Hung}, T., {Cenko}, S.~B., {Roth}, N., {et~al.} 2019, \apj, 879, 119, \dodoi{10.3847/1538-4357/ab24de}

\bibitem[{{Hung} {et~al.}(2021){Hung}, {Foley}, {Veilleux}, {Cenko}, {Dai}, {Auchettl}, {Brink}, {Dimitriadis}, {Filippenko}, {Gezari}, {Holoien}, {Kilpatrick}, {Mockler}, {Piro}, {Ramirez-Ruiz}, {Rojas-Bravo}, {Siebert}, {van Velzen}, \& {Zheng}}]{Hung2021}
{Hung}, T., {Foley}, R.~J., {Veilleux}, S., {et~al.} 2021, \apj, 917, 9, \dodoi{10.3847/1538-4357/abf4c3}

\bibitem[{{Jiang} {et~al.}(2014){Jiang}, {Stone}, \& {Davis}}]{Jiang2014}
{Jiang}, Y.-F., {Stone}, J.~M., \& {Davis}, S.~W. 2014, \apj, 796, 106, \dodoi{10.1088/0004-637X/796/2/106}

\bibitem[{{Jin} {et~al.}(2025){Jin}, {Li}, {Jiang}, {Dai}, {Cheng}, {Zhu}, {Yang}, {Rau}, {Baldini}, {Wang}, {Zhou}, {Yuan}, {Zhang}, {Shu}, {Shen}, {Wang}, {Wen}, {Wu}, {Wang}, {Thomsen}, {Zhang}, {Zhang}, {Coleiro}, {Eyles-Ferris}, {Fang}, {Ho}, {Hu}, {Jin}, {Li}, {Liu}, {Liu}, {Liu}, {Liu}, {Lu}, {Merloni}, {Qiao}, {Saxton}, {Soria}, {Wang}, {Xue}, {Yang}, {Zhang}, {Zhang}, {Cai}, {Chen}, {Chen}, {Chen}, {Chen}, {Chen}, {Chen}, {Chen}, {Cordier}, {Cui}, {Cui}, {Dai}, {Ding}, {Fan}, {Fan}, {Feng}, {Garcia}, {Guan}, {Han}, {Hou}, {Hu}, {Huang}, {Huo}, {Jia}, {Jia}, {Jiang}, {Jin}, {Kong}, {Kuulkers}, {Lei}, {Li}, {Li}, {Li}, {Li}, {Li}, {Li}, {Lian}, {Ling}, {Liu}, {Liu}, {Liu}, {Liu}, {Liu}, {Lu}, {Luo}, {Ma}, {Mao}, {Mu}, {Nandra}, {O'Brien}, {Pan}, {Pan}, {Qin}, {Rea}, {Sanders}, {Song}, {Sun}, {Sun}, {Sun}, {Tan}, {Tang}, {Tao}, {Wang}, {Wang}, {Wang}, {Wang}, {Wang}, {Wang}, {Wang}, {Wu}, {Wu}, {Xu}, {Xu}, {Xu}, {Xu}, {Xu}, {Xue}, {Xue}, {Xue}, {Yan}, {Yang}, {Yang}, {Zhang}, {Zhang}, {Zhang}, {Zhang},
  {Zhang}, {Zhang}, {Zhang}, {Zhao}, {Zhao}, {Zhao}, {Zhao}, {Zheng}, {Zhu}, {Zhu}, {Zhu}, \& {Zou}}]{Jin2025}
{Jin}, C.~C., {Li}, D.~Y., {Jiang}, N., {et~al.} 2025, arXiv e-prints, arXiv:2501.09580, \dodoi{10.48550/arXiv.2501.09580}

\bibitem[{{Kaastra}(2017)}]{Kaastra2017}
{Kaastra}, J.~S. 2017, \aap, 605, A51, \dodoi{10.1051/0004-6361/201629319}

\bibitem[{{Kaastra} \& {Bleeker}(2016)}]{Kaastra2016}
{Kaastra}, J.~S., \& {Bleeker}, J.~A.~M. 2016, \aap, 587, A151, \dodoi{10.1051/0004-6361/201527395}

\bibitem[{{Kaastra} {et~al.}(1996){Kaastra}, {Mewe}, \& {Nieuwenhuijzen}}]{Kaastra1996}
{Kaastra}, J.~S., {Mewe}, R., \& {Nieuwenhuijzen}, H. 1996, in UV and X-ray Spectroscopy of Astrophysical and Laboratory Plasmas, ed. K.~{Yamashita} \& T.~{Watanabe}, 411--414

\bibitem[{{Kara} {et~al.}(2018){Kara}, {Dai}, {Reynolds}, \& {Kallman}}]{Kara2018}
{Kara}, E., {Dai}, L., {Reynolds}, C.~S., \& {Kallman}, T. 2018, \mnras, 474, 3593, \dodoi{10.1093/mnras/stx3004}

\bibitem[{{Kosec} {et~al.}(2023){Kosec}, {Pasham}, {Kara}, \& {Tombesi}}]{Kosec2023}
{Kosec}, P., {Pasham}, D., {Kara}, E., \& {Tombesi}, F. 2023, \apj, 954, 170, \dodoi{10.3847/1538-4357/aced87}

\bibitem[{{Kosec} {et~al.}(2025){Kosec}, {Kara}, {Brenneman}, {Chakraborty}, {Giustini}, {Miniutti}, {Pinto}, {Rogantini}, {Arcodia}, {Middleton}, \& {Sacchi}}]{Kosec2025}
{Kosec}, P., {Kara}, E., {Brenneman}, L., {et~al.} 2025, \apj, 978, 10, \dodoi{10.3847/1538-4357/ad9249}

\bibitem[{{Krolik} \& {Piran}(2012)}]{Krolik2012}
{Krolik}, J.~H., \& {Piran}, T. 2012, \apj, 749, 92, \dodoi{10.1088/0004-637X/749/1/92}

\bibitem[{{Lin} {et~al.}(2015){Lin}, {Maksym}, {Irwin}, {Komossa}, {Webb}, {Godet}, {Barret}, {Grupe}, \& {Gwyn}}]{Lin2015ApJ...811...43L}
{Lin}, D., {Maksym}, P.~W., {Irwin}, J.~A., {et~al.} 2015, \apj, 811, 43, \dodoi{10.1088/0004-637X/811/1/43}

\bibitem[{{Lin} {et~al.}(2018){Lin}, {Strader}, {Carrasco}, {Page}, {Romanowsky}, {Homan}, {Irwin}, {Remillard}, {Godet}, {Webb}, {Baumgardt}, {Wijnands}, {Barret}, {Duc}, {Brodie}, \& {Gwyn}}]{Lin2018}
{Lin}, D., {Strader}, J., {Carrasco}, E.~R., {et~al.} 2018, Nature Astronomy, 2, 656, \dodoi{10.1038/s41550-018-0493-1}

\bibitem[{{Lodato} \& {Rossi}(2011)}]{Lodato2011}
{Lodato}, G., \& {Rossi}, E.~M. 2011, \mnras, 410, 359, \dodoi{10.1111/j.1365-2966.2010.17448.x}

\bibitem[{{Lodders} {et~al.}(2009){Lodders}, {Palme}, \& {Gail}}]{Lodders2009}
{Lodders}, K., {Palme}, H., \& {Gail}, H.~P. 2009, Landolt B{\"o}rnstein, 4B, 712, \dodoi{10.1007/978-3-540-88055-4_34}

\bibitem[{{Loeb} \& {Ulmer}(1997)}]{Loeb1997}
{Loeb}, A., \& {Ulmer}, A. 1997, \apj, 489, 573, \dodoi{10.1086/304814}

\bibitem[{{Lu} \& {Kumar}(2018)}]{Lu2018ApJ}
{Lu}, W., \& {Kumar}, P. 2018, \apj, 865, 128, \dodoi{10.3847/1538-4357/aad54a}

\bibitem[{{Makrygianni} {et~al.}(2023){Makrygianni}, {Trakhtenbrot}, {Arcavi}, {Ricci}, {Lam}, {Horesh}, {Sfaradi}, {Bostroem}, {Hosseinzadeh}, {Howell}, {Pellegrino}, {Fender}, {Green}, {Williams}, \& {Bright}}]{Makrygianni2023}
{Makrygianni}, L., {Trakhtenbrot}, B., {Arcavi}, I., {et~al.} 2023, \apj, 953, 32, \dodoi{10.3847/1538-4357/ace1ee}

\bibitem[{{Matzeu} {et~al.}(2017){Matzeu}, {Reeves}, {Braito}, {Nardini}, {McLaughlin}, {Lobban}, {Tombesi}, \& {Costa}}]{Matzeu2017}
{Matzeu}, G.~A., {Reeves}, J.~N., {Braito}, V., {et~al.} 2017, \mnras, 472, L15, \dodoi{10.1093/mnrasl/slx129}

\bibitem[{{McKinney} {et~al.}(2014){McKinney}, {Tchekhovskoy}, {Sadowski}, \& {Narayan}}]{McKinney2014}
{McKinney}, J.~C., {Tchekhovskoy}, A., {Sadowski}, A., \& {Narayan}, R. 2014, \mnras, 441, 3177, \dodoi{10.1093/mnras/stu762}

\bibitem[{{Mehdipour} {et~al.}(2016){Mehdipour}, {Kaastra}, \& {Kallman}}]{Mehdipour2016}
{Mehdipour}, M., {Kaastra}, J.~S., \& {Kallman}, T. 2016, \aap, 596, A65, \dodoi{10.1051/0004-6361/201628721}

\bibitem[{{Metzger} \& {Stone}(2016)}]{Metzger2016}
{Metzger}, B.~D., \& {Stone}, N.~C. 2016, \mnras, 461, 948, \dodoi{10.1093/mnras/stw1394}

\bibitem[{{Miller} {et~al.}(2015){Miller}, {Kaastra}, {Miller}, {Reynolds}, {Brown}, {Cenko}, {Drake}, {Gezari}, {Guillochon}, {Gultekin}, {Irwin}, {Levan}, {Maitra}, {Maksym}, {Mushotzky}, {O'Brien}, {Paerels}, {de Plaa}, {Ramirez-Ruiz}, {Strohmayer}, \& {Tanvir}}]{Miller2015}
{Miller}, J.~M., {Kaastra}, J.~S., {Miller}, M.~C., {et~al.} 2015, \nat, 526, 542, \dodoi{10.1038/nature15708}

\bibitem[{{Mitsuda} {et~al.}(1984){Mitsuda}, {Inoue}, {Koyama}, {Makishima}, {Matsuoka}, {Ogawara}, {Shibazaki}, {Suzuki}, {Tanaka}, \& {Hirano}}]{Mitsuda1984}
{Mitsuda}, K., {Inoue}, H., {Koyama}, K., {et~al.} 1984, \pasj, 36, 741

\bibitem[{{Nardini} {et~al.}(2015){Nardini}, {Reeves}, {Gofford}, {Harrison}, {Risaliti}, {Braito}, {Costa}, {Matzeu}, {Walton}, {Behar}, {Boggs}, {Christensen}, {Craig}, {Hailey}, {Matt}, {Miller}, {O'Brien}, {Stern}, {Turner}, \& {Ward}}]{Nardini2015}
{Nardini}, E., {Reeves}, J.~N., {Gofford}, J., {et~al.} 2015, Science, 347, 860, \dodoi{10.1126/science.1259202}

\bibitem[{{Ohsuga} {et~al.}(2005){Ohsuga}, {Mori}, {Nakamoto}, \& {Mineshige}}]{Ohsuga2005}
{Ohsuga}, K., {Mori}, M., {Nakamoto}, T., \& {Mineshige}, S. 2005, \apj, 628, 368, \dodoi{10.1086/430728}

\bibitem[{{Pasham} {et~al.}(2024){Pasham}, {Tombesi}, {Sukov{\'a}}, {Zaja{\v{c}}ek}, {Rakshit}, {Coughlin}, {Kosec}, {Karas}, {Masterson}, {Mummery}, {Holoien}, {Guolo}, {Hinkle}, {Ripperda}, {Witzany}, {Shappee}, {Kara}, {Horesh}, {van Velzen}, {Sfaradi}, {Kaplan}, {Burger}, {Murphy}, {Remillard}, {Steiner}, {Wevers}, {Arcodia}, {Buchner}, {Merloni}, {Malyali}, {Fabian}, {Fausnaugh}, {Daylan}, {Altamirano}, {Payne}, \& {Ferraraa}}]{Pasham2024}
{Pasham}, D.~R., {Tombesi}, F., {Sukov{\'a}}, P., {et~al.} 2024, Science Advances, 10, eadj8898, \dodoi{10.1126/sciadv.adj8898}

\bibitem[{{Payne} {et~al.}(2023){Payne}, {Auchettl}, {Shappee}, {Kochanek}, {Boyd}, {Holoien}, {Fausnaugh}, {Ashall}, {Hinkle}, {Vallely}, {Stanek}, \& {Thompson}}]{Payne2023}
{Payne}, A.~V., {Auchettl}, K., {Shappee}, B.~J., {et~al.} 2023, \apj, 951, 134, \dodoi{10.3847/1538-4357/acd455}

\bibitem[{{Phinney}(1989)}]{Phinney1989}
{Phinney}, E.~S. 1989, in IAU Symposium, Vol. 136, The Center of the Galaxy, ed. M.~{Morris}, 543

\bibitem[{{Pinto} {et~al.}(2018){Pinto}, {Alston}, {Parker}, {Fabian}, {Gallo}, {Buisson}, {Walton}, {Kara}, {Jiang}, {Lohfink}, \& {Reynolds}}]{Pinto2018}
{Pinto}, C., {Alston}, W., {Parker}, M.~L., {et~al.} 2018, \mnras, 476, 1021, \dodoi{10.1093/mnras/sty231}

\bibitem[{{Pinto} {et~al.}(2020){Pinto}, {Mehdipour}, {Walton}, {Middleton}, {Roberts}, {Fabian}, {Guainazzi}, {Soria}, {Kosec}, \& {Ness}}]{Pinto2020MNRAS.491.5702P}
{Pinto}, C., {Mehdipour}, M., {Walton}, D.~J., {et~al.} 2020, \mnras, 491, 5702, \dodoi{10.1093/mnras/stz3392}

\bibitem[{{Qiao} {et~al.}(2025){Qiao}, {Wu}, {Lin}, {Guo}, {Liu}, {Guo}, {Jin}, \& {Jiang}}]{Qiao2025}
{Qiao}, E., {Wu}, Y., {Lin}, Y., {et~al.} 2025, \mnras, 539, 3473, \dodoi{10.1093/mnras/staf719}

\bibitem[{{Rees}(1988)}]{Rees1988Nat}
{Rees}, M.~J. 1988, \nat, 333, 523, \dodoi{10.1038/333523a0}

\bibitem[{{{\'S}niegowska} {et~al.}(2025){{\'S}niegowska}, {Trakhtenbrot}, {Makrygianni}, {Arcavi}, {Ricci}, {Faris}, {Palit}, {Howell}, {Newsome}, {Farah}, {McCully}, {Padilla-Gonzalez}, \& {Terreran}}]{Sniegowska2025}
{{\'S}niegowska}, M., {Trakhtenbrot}, B., {Makrygianni}, L., {et~al.} 2025, arXiv e-prints, arXiv:2505.00083, \dodoi{10.48550/arXiv.2505.00083}

\bibitem[{{Strubbe} \& {Quataert}(2009)}]{Strubbe2009}
{Strubbe}, L.~E., \& {Quataert}, E. 2009, \mnras, 400, 2070, \dodoi{10.1111/j.1365-2966.2009.15599.x}

\bibitem[{{Svirski} {et~al.}(2017){Svirski}, {Piran}, \& {Krolik}}]{Svirski2017}
{Svirski}, G., {Piran}, T., \& {Krolik}, J. 2017, \mnras, 467, 1426, \dodoi{10.1093/mnras/stx117}

\bibitem[{{Tashiro}(2022)}]{Tashiro2022}
{Tashiro}, M.~S. 2022, International Journal of Modern Physics D, 31, 2230001, \dodoi{10.1142/S0218271822300014}

\bibitem[{{Thomsen} {et~al.}(2022){Thomsen}, {Kwan}, {Dai}, {Wu}, {Roth}, \& {Ramirez-Ruiz}}]{Thomsen2022}
{Thomsen}, L.~L., {Kwan}, T.~M., {Dai}, L., {et~al.} 2022, \apjl, 937, L28, \dodoi{10.3847/2041-8213/ac911f}

\bibitem[{{Trakhtenbrot} {et~al.}(2019){Trakhtenbrot}, {Arcavi}, {Ricci}, {Tacchella}, {Stern}, {Netzer}, {Jonker}, {Horesh}, {Mej{\'\i}a-Restrepo}, {Hosseinzadeh}, {Hallefors}, {Howell}, {McCully}, {Balokovi{\'c}}, {Heida}, {Kamraj}, {Lansbury}, {Wyrzykowski}, {Gromadzki}, {Hamanowicz}, {Cenko}, {Sand}, {Hsiao}, {Phillips}, {Diamond}, {Kara}, {Gendreau}, {Arzoumanian}, \& {Remillard}}]{Trakhtenbrot2019}
{Trakhtenbrot}, B., {Arcavi}, I., {Ricci}, C., {et~al.} 2019, Nature Astronomy, 3, 242, \dodoi{10.1038/s41550-018-0661-3}

\bibitem[{{van Velzen} {et~al.}(2019){van Velzen}, {Stone}, {Metzger}, {Gezari}, {Brown}, \& {Fruchter}}]{vanVelzen2019}
{van Velzen}, S., {Stone}, N.~C., {Metzger}, B.~D., {et~al.} 2019, \apj, 878, 82, \dodoi{10.3847/1538-4357/ab1844}

\bibitem[{{Verner} {et~al.}(1996){Verner}, {Ferland}, {Korista}, \& {Yakovlev}}]{Verner1996}
{Verner}, D.~A., {Ferland}, G.~J., {Korista}, K.~T., \& {Yakovlev}, D.~G. 1996, \apj, 465, 487, \dodoi{10.1086/177435}

\bibitem[{{Wang} {et~al.}(2022){Wang}, {Bu}, \& {Yuan}}]{Wang2022MNRAS.513.5818W}
{Wang}, W., {Bu}, D.-F., \& {Yuan}, F. 2022, \mnras, 513, 5818, \dodoi{10.1093/mnras/stac1348}

\bibitem[{{Wang} {et~al.}(2021){Wang}, {Ji}, {Garc{\'\i}a}, {Dauser}, {M{\'e}ndez}, {Mao}, {Tao}, {Altamirano}, {Maggi}, {Zhang}, {Ge}, {Zhang}, {Qu}, {Zhang}, {Ma}, {Lu}, {Li}, {Huang}, {Zheng}, {Chang}, {Tuo}, {Song}, {Xu}, {Chen}, {Liu}, {Bu}, {Cai}, {Cao}, {Chen}, {Chen}, {Chen}, {Cui}, {Du}, {Gao}, {Gu}, {Guan}, {Guo}, {Han}, {Huo}, {Jia}, {Jiang}, {Jin}, {Kong}, {Li}, {Li}, {Li}, {Li}, {Li}, {Li}, {Li}, {Li}, {Liang}, {Liao}, {Liu}, {Liu}, {Lu}, {Luo}, {Luo}, {Meng}, {Nang}, {Nie}, {Ou}, {Sai}, {Shang}, {Song}, {Sun}, {Tan}, {Wang}, {Wang}, {Wang}, {Wen}, {Wu}, {Wu}, {Wu}, {Xiao}, {Xiao}, {Xiong}, {Yang}, {Yang}, {Yi}, {Yin}, {You}, {Zhang}, {Zhang}, {Zhang}, {Zhang}, {Zhang}, {Zhang}, {Zhao}, {Zhao}, \& {Zhou}}]{Wang2021ApJ...906...11W}
{Wang}, Y., {Ji}, L., {Garc{\'\i}a}, J.~A., {et~al.} 2021, \apj, 906, 11, \dodoi{10.3847/1538-4357/abc55e}

\bibitem[{{Wevers} {et~al.}(2023){Wevers}, {Guolo}, {Pasham}, {Coughlin}, {Tombesi}, {Yao}, \& {Gezari}}]{Wevers2023}
{Wevers}, T., {Guolo}, M., {Pasham}, D.~R., {et~al.} 2023, arXiv e-prints, arXiv:2311.09371, \dodoi{10.48550/arXiv.2311.09371}

\bibitem[{{Wilms} {et~al.}(2000){Wilms}, {Allen}, \& {McCray}}]{Wilms2000}
{Wilms}, J., {Allen}, A., \& {McCray}, R. 2000, \apj, 542, 914, \dodoi{10.1086/317016}

\bibitem[{{Xiang} {et~al.}(2024){Xiang}, {Miller}, {Zoghbi}, {Reynolds}, {Bogensberger}, {Dai}, {Draghis}, {Drake}, {Godet}, {Irwin}, {Miller}, {Mockler}, {Saxton}, \& {Webb}}]{Xiang2024}
{Xiang}, X., {Miller}, J.~M., {Zoghbi}, A., {et~al.} 2024, \apj, 972, 106, \dodoi{10.3847/1538-4357/ad6002}

\bibitem[{{Xu} {et~al.}(2024){Xu}, {Pinto}, {Rogantini}, {Barret}, {Bianchi}, {Guainazzi}, {Ebrero}, {Alston}, {Kara}, \& {Cusumano}}]{Xu2024}
{Xu}, Y., {Pinto}, C., {Rogantini}, D., {et~al.} 2024, \aap, 687, A179, \dodoi{10.1051/0004-6361/202349001}

\bibitem[{{Yang} {et~al.}(2023){Yang}, {Yuan}, {Kwan}, \& {Dai}}]{Yang2023}
{Yang}, H., {Yuan}, F., {Kwan}, T., \& {Dai}, L. 2023, \mnras, 523, 208, \dodoi{10.1093/mnras/stad1444}

\end{thebibliography}
\bibliographystyle{aasjournal}

\appendix

\renewcommand{\thefigure}{\hbAppendixPrefix\arabic{figure}}
\numberwithin{figure}{section}
\renewcommand{\thetable}{\hbAppendixPrefix\arabic{table}}
\numberwithin{table}{section}

\section{Identification of Absorption Features}\label{sec:abs_feature}

\begin{figure}[!h]
 \centering
   \includegraphics[width=1\linewidth]{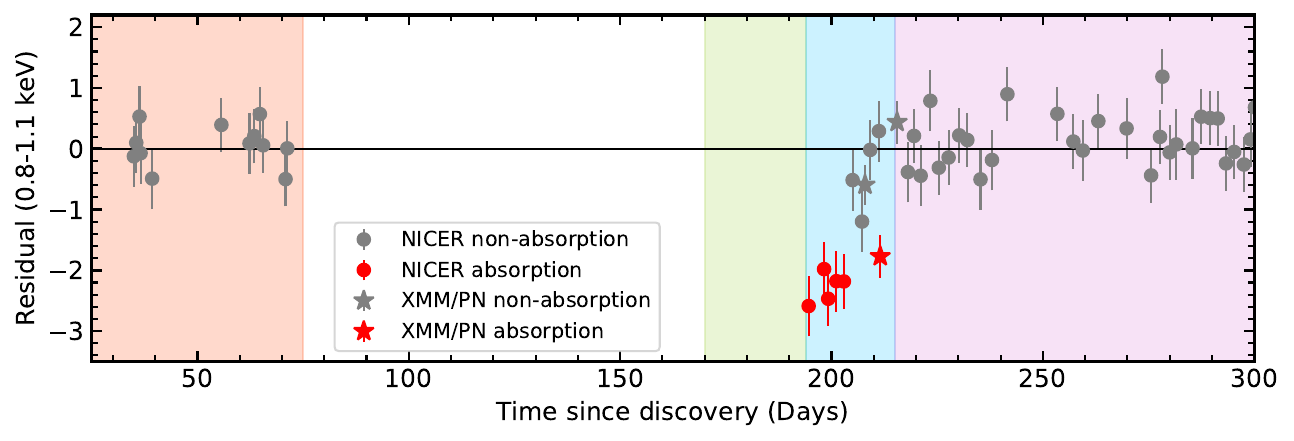}
   \caption{The average residual of \nicer (circles) and \xmm/PN (stars) continuum spectrum within 0.8--1.1\,keV. Red symbols represent observations with detected absorption features, while gray symbols indicate those without. The color code of shaded regions is same as Fig.~\ref{fig:lc}.}
   \label{fig:NICER_res}
\end{figure}

In this section, we applied a method to quantify the absorption features in individual \nicer and \xmm/PN observations. We calculated the average residual, defined as \texttt{(Data-Model)/Error}, within the 0.8–1.1\,keV band of the best-fitting continuum spectrum. We found that only the six observations in epoch (3) show significant ($>3\sigma$) deviations from the best-fitting continuum (see Fig.~\ref{fig:NICER_res}). This suggests that the absorption feature is absent in the other observations. Additionally, we report the minimum, median, and maximum total counts for individual observations in each epoch in Table~\ref{tab:nicer_epoch}. Note that no \nicer observation was performed during epoch (3). As shown in the table, the total number of counts in each epoch is comparable. Therefore, we conclude that the non-detections in individual \nicer observations during epochs (1) and (4) are not due to low SNRs.

\begin{deluxetable}{lccccccc}[!h]
\tablecaption{Epoch log of \nicer observations.} 
\label{tab:nicer_epoch}
\tablewidth{0pt}
\tablehead{
\colhead{Epoch} & \colhead{Start Time} & \colhead{End Time} & \colhead{Phase$^{a}$} & \colhead{C$_{\rm min}^{b}$} & \colhead{C$_{\rm median}^{b}$} & \colhead{C$_{\rm max}^{b}$} &\\[-5pt]
 & (MJD) & (MJD) & (days) & (counts) & (counts) & (counts) & }
\startdata
Epoch (1) & 60344.9685 &  60381.5262 & 35.0-71.5 & 830 & 2266 & 8786\\
Epoch (3) & 60504.3363 &  60521.3228 & 194.3-211.3 & 993 & 3612 & 9793\\
Epoch (4) & 60528.0296 &  60611.3076 & 218.0-301.3 & 376 & 2426 & 10485\\
\enddata
\tablecomments{
$^{a}$: Days since discovery (MJD 60310).\\
$^{b}$: C$_{\rm min}$, C$_{\rm median}$, and C$_{\rm max}$ denote the minimum, median, and maximum total counts in the corresponding epoch.
}
\end{deluxetable}

\clearpage

\section{Photoionization spectroscopy parameters} \label{sec:SPEX_spe}
\begin{figure}[!h]
 \centering
   \includegraphics[width=1\linewidth]{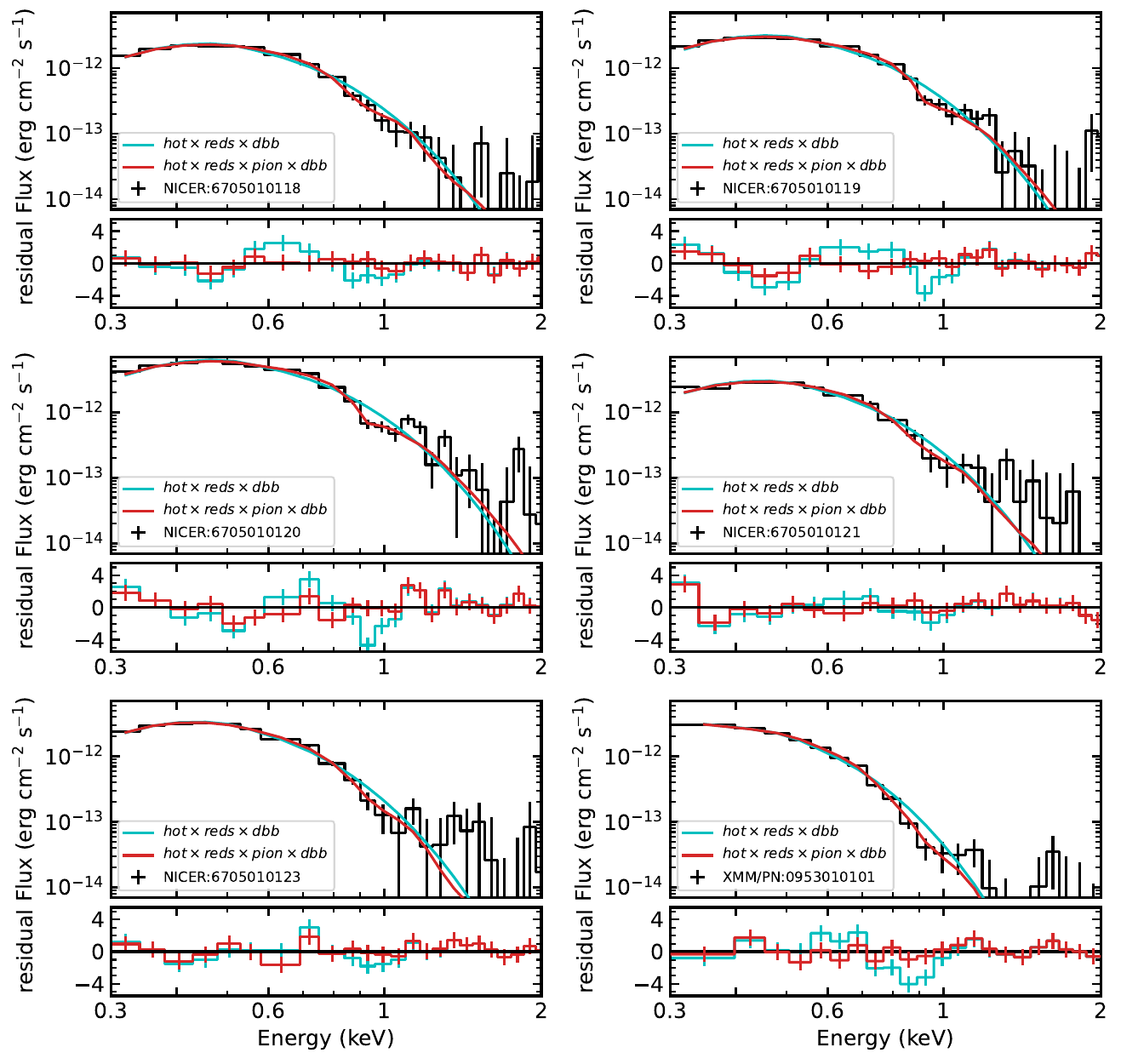}
   \caption{Example of the {\sc SPEX} spectral fitting model. The cyan and red curves, along with their corresponding residuals, represent the continuum model and the continuum model with the \pion component, respectively.}
   \label{fig:SPEX_spe}
\end{figure}

In this section, we present the results and spectral fitting plots as described in Sect.~\ref{sec:Result}. Fig.\ref{fig:SPEX_spe} shows the comparison of spectra and residuals with and without the inclusion of the \texttt{pion} model in the continuum.

\begin{deluxetable}{lccccccccc}
\tablecaption{Best-fitting parameters for observations in which absorption features were detected. The applied models are \texttt{tbabs}$\times$\texttt{zashift}$\times$(\texttt{gaussian}$+$\texttt{diskbb}) in {\sc xspec} and \texttt{hot}$\times$\texttt{reds}$\times$\texttt{pion}$\times$\texttt{dbb} in {\sc SPEX}.}
\tablewidth{0pt}
\tablehead{
  \colhead{Model} & \colhead{Parameter} 
  & \multicolumn{5}{c}{\nicer ObsID} & \multicolumn{1}{c}{\xmm ObsID}\\
  & & 6705010118 & 6705010119 & 6705010120 & 6705010121 & 6705010123 & 0953010101 \\
  \hline
  & Phase$^{a}$ (days) & 194.3 & 198.0 & 199.1 & 201.1 & 203.0 & 211.6
}
\startdata
\hline
\multicolumn{8}{c}{\sc xspec}\\
\hline
  tbabs & $N_{\rm H}$ ($10^{22}$\,cm$^{-2}$) & \multicolumn{6}{c}{0.047} \\ 
\hline
  zashift & z & \multicolumn{6}{c}{0.027}  \\ 
\hline
  diskbb 
  & $T_{in}$ (keV) & $0.128^{+0.004}_{-0.003}$ & $0.126\pm0.002$ & $0.134^{+0.001}_{-0.002}$ & $0.135^{+0.020}_{-0.009}$ & $0.113^{+0.003}_{-0.002}$ & $0.091^{+0.003}_{-0.002}$\\ 
  & norm ($10^{3}$) & $1.6^{+0.1}_{-0.2}$ & $2.26^{+0.08}_{-0.20}$ & $3.4^{+0.1}_{-0.4}$ & $1.5^{+0.1}_{-0.4}$ & $4.3^{+0.2}_{-0.5}$ & $14.0^{+2.0}_{-2.4}$ \\
\hline
  gauss  
  & $E_{\rm line}$ (keV) &  $0.90^{+0.03}_{-0.05}$ & $0.95^{+0.01}_{-0.02}$ & $0.94\pm0.01$ & $0.8\pm0.1$ & $0.91^{+0.03}_{-0.04}$ & $0.78^{+0.04}_{-0.07}$  \\ 
  & $E_\sigma$ (keV) & $0.13^{+0.04}_{-0.03}$ & $0.07\pm0.02$ & $0.06^{+0.01}_{-0.02}$ & $0.18^{+0.03}_{-0.06}$ & $0.11^{+0.02}_{-0.03}$ & $0.12^{+0.04}_{-0.03}$ \\ 
  & $-E_{\rm norm}$$^{b}$ ($10^{-5}$) & $7.8^{+4.2}_{-2.4}$ & $5.1^{+1.4}_{-1.1}$ & $13.8^{+3.4}_{-2.0}$ & $27.3^{+20.0}_{-15.6}$ & $6.9^{+2.9}_{-2.1}$  & $6.6^{+6.4}_{-2.3}$ \\
  & \dstat$^{c}$ & $34.7$ & $41.9$ & $53.6$ & $24.1$ & $34.8$ & $66.2$ \\
\hline
  C-stat/$\nu$ & & $116.41/107$ & $125.85/126$ & $103.24/101$ & $92.02/103$ & $115.18/130$ & $136.58/137$\\
\hline
\hline
\multicolumn{8}{c}{\sc SPEX}\\
\hline
  hot    
  & $N_{\rm H}$ ($10^{22}$\,cm$^{-2}$) & \multicolumn{6}{c}{0.047} \\ 
\hline
  reds   
  & z & \multicolumn{6}{c}{0.027} \\ 
\hline
  dbb$^{d}$    
  & $T_{in}$ (keV) & $0.244^{+0.010}_{-0.005}$ & $0.240^{+0.003}_{-0.005}$ & $0.252^{+0.004}_{-0.003}$ & $0.229\pm0.007$ & $0.215\pm0.005$ & $0.174^{+0.002}_{-0.003}$ \\
   & norm & $14^{+1}_{-2}$ & $21^{+2}_{-1}$ & $33^{+1}_{-2}$ & $25^{+2}_{-4}$ & $36^{+5}_{-3}$ & $137^{+14}_{-12}$ \\ 
\hline
  pion
  & $N_{\rm H}$ ($10^{22}$\,cm$^{-2}$) & $3.2^{+2.0}_{-0.7}$ & $0.8\pm0.2$ & $0.8^{+0.2}_{-0.1}$ & $2.2^{+1.0}_{-0.9}$ & $2.9^{+2.0}_{-0.9}$ & $2.8\pm1.0$ \\ 
  & $\log\xi$ ($\rm erg\;cm\;s^{-1}$) &  $2.6\pm0.2$ & $2.1\pm0.1$ & $2.0\pm0.1$ & $2.6\pm0.2$ & $2.9\pm0.3$ & $2.9\pm0.1$ \\ 
  & $v_{\rm rms}$ ($\rm km\;s^{-1}$) & $129^{+146}_{-85}$ & $143^{+241}_{-86}$ & $161^{+223}_{-77}$ & $292^{+1061}_{-206}$ & $80^{+868}_{-58}$ & $21^{+18}_{-8}$ \\ 
  & $v_{\rm out}/c$ &  $-0.015^{+0.012}_{-0.119}$ & $-0.186^{+0.012}_{-0.029}$ & $-0.190^{+0.021}_{-0.010}$ & $-0.036^{+0.013}_{-0.304}$ & $-0.043^{+0.011}_{-0.270}$ & $-0.0097^{+0.0008}_{-0.0005}$ \\ 
  & \dstat$^{c}$ &  $26.8$ & $45.0$ & $45.3$ & $10.3$ & $16.1$ & $61.1$ \\ 
\hline
  C-stat/$\nu$ & & $14.20/18$ & $25.28/19$ & $48.30/19$ & $27.17/17$ & $29.09/18$ & $531.70/320$\\
\enddata
\tablecomments{
$^{a}$: Days since discovery (MJD 60310).\\
$^{b}$: The norm of the Gaussian parameter was taken as its absolute value to clearly represent the line strength.\\
$^{c}$: \dstat stands for the improvement in the inclusion of this component.\\
$^{d}$: The \texttt{dbb} model in {\sc SPEX} has a different definition compared to the \texttt{diskbb} model in {\sc xspec}, which is approximately greater by a factor of 2.}
\label{tab:spe_result}
\end{deluxetable}

\begin{deluxetable}{lccccccc}[!ht]
\tablecaption{Best-fitting parameters for \swift in four epochs and two individual \xmm/PN spectra. The applied model is \texttt{tbabs}$\times$\texttt{zashift}$\times$(\texttt{gaussian}$+$\texttt{diskbb}) in {\sc xspec}. For spectra exhibiting a significant hard excess, an additional \texttt{powerlaw} component was included.}
\tablewidth{0pt}
\tablehead{
  \colhead{Model} & \colhead{Parameter} 
  & \multicolumn{4}{c}{\swift} & \multicolumn{2}{c}{\xmm Obsid} \\
  & & epoch (1) & epoch (2) & epoch (3) & epoch (4) & 0953010201 & 0953010301\\
  \hline
  & Phase$^{a}$ (days) & 29.7--74.2 & 172.0--193.4 & 196.1--214.7 & 220.1--301.9 & 208.0 & 215.6
}
\startdata
\hline
\multicolumn{8}{c}{\sc xspec}\\
\hline
  tbabs & $N_{\rm H}$ ($10^{22}$\,cm$^{-2}$) & \multicolumn{5}{c}{0.047} \\ 
\hline
  zashift & z & \multicolumn{5}{c}{0.027}  \\ 
\hline
  diskbb 
  & $T_{in}$ (keV) & $0.100^{+0.003}_{-0.002}$ & $0.107\pm0.005$ & $0.131\pm0.005$ & $0.12\pm0.01$ & $0.095^{+0.006}_{-0.005}$ & $0.074^{+0.003}_{-0.004}$  \\ 
  & norm ($10^{3}$) & $4.4\pm0.7$ & $3.7^{+1.1}_{-1.0}$ & $1.1\pm0.2$ & $0.3\pm0.1$ & $6.0^{+3.4}_{-2.1}$ & $4.7^{+1.8}_{-1.2}$  \\ 
\hline
  powerlaw 
  & $\Gamma$ & - & $3.3^{+0.9}_{-1.0}$ & - & $2.7\pm0.3$ & $5.7\pm0.3$ & $2.7^{+0.9}_{-0.8}$   \\
  & norm ($10^{-5}$) & - & $5.8^{+3.6}_{-2.6}$ & - & $5.4^{+1.0}_{-1.2}$ & $5.1^{+0.7}_{-0.8}$ & $0.9^{+0.3}_{-0.2}$  \\
\hline
  gauss  
  & $E_{\rm line}$ (keV) &  $0.93$(f) & $0.93^{+0.04}_{-0.03}$ & $0.96^{+0.02}_{-0.03}$  &  $0.78$(f) & $0.78$(f) & $0.78$(f) \\ 
  & $E_\sigma$ (keV) & $0.05$(f) & $0.05$(f) & $0.05$(f) & $0.12$(f) & $0.12$(f) & $0.12$(f) \\ 
  & $-E_{\rm norm}$$^{b}$ ($10^{-5}$) & $<1.9$ & $3.4^{+1.4}_{-1.3}$ & $5.0\pm1.2$ & $<4.2$ & $3.7\pm1.2$ & $<1.5$ \\ 
  & \dstat$^{c}$ & $0.0$ & $7.1$ & $13.2$ & $0.0$ & $8.9$ & $2.1$\\ 
\hline
  C-stat/$\nu$ & & $10.59/10$ & $3.82/10$ & $14.78/10$ & $27.29/22$ & $81.42/96$ & $93.65/88$ \\
\enddata
\tablecomments{
$^{a}$: Days since discovery (MJD 60310).\\
$^{b}$: The $E_{\rm norm}$ of the Gaussian parameter was taken as its absolute value to clearly represent the line strength.\\
$^{c}$: \dstat stands for the improvement in the inclusion of this component.\\
(f) represent that the parameter was fixed during the fitting.}
\label{tab:swift_spe}
\end{deluxetable}

\end{document}